\documentstyle[epsf,times]{mn}

\title[The ROSAT Brightest Cluster Sample (BCS) --- I. The sample]
      {The ROSAT Brightest Cluster Sample (BCS) --- I. The compilation of the 
       sample and the cluster log N-log S distribution}
\author[H.\ Ebeling et al.]
       {\parbox{\textwidth}{H.\ Ebeling$^{1,2,3}$,    
        A.C.\ Edge$^2$, H.\ B\"ohringer$^1$, S.W.\ Allen$^2$, 
        C.S.\ Crawford$^2$, A.C.\ Fabian$^2$, W.\ Voges$^1$, 
        J.P.\ Huchra$^4$ }\\ \\
        $^1$ Max-Planck-Institut f\"ur extraterrestrische Physik, 
             Giessenbachstr., D-85740 Garching, Germany\\
 	$^2$ Institute of Astronomy, Madingley Road, Cambridge CB3\,0HA, UK\\
        $^3$ Institute for Astronomy, 2680 Woodlawn Dr, Honolulu HI 96822, USA;
             email: {\em ebeling{@}ifa.hawaii.edu}\\
	$^4$ Harvard-Smithsonian Center for Astrophysics, 60 Garden Street, 
             Cambridge, MA 02138, USA} 

\date{To appear in December 1998 in Volume 301, pp881--914}
\begin{document}

\maketitle

\begin{abstract}
We present a 90 per cent flux-complete sample of the 201 X-ray
brightest clusters of galaxies in the northern hemisphere ($\delta
\geq 0^{\circ}$), at high Galactic latitudes ($|b| \geq 20^{\circ}$),
with measured redshifts $z \leq 0.3$ and fluxes higher than $4.4\times
10^{-12}$ erg cm$^{-2}$ s$^{-1}$ in the 0.1--2.4 keV band.  The
sample, called the ROSAT Brightest Cluster Sample (BCS), is selected
from ROSAT All-Sky Survey data and is the largest X-ray selected
cluster sample compiled to date. In addition to Abell clusters, which
form the bulk of the sample, the BCS also contains the X-ray brightest
Zwicky clusters and other clusters selected from their X-ray
properties alone. Effort has been made to ensure the highest possible
completeness of the sample and the smallest possible contamination by
non-cluster X-ray sources.  X-ray fluxes are computed using an
algorithm tailored for the detection and characterization of X-ray
emission from galaxy clusters. These fluxes are accurate to better
than 15 per cent (mean $1\sigma$ error).

We find the cumulative $\log N-\log S$ distribution of clusters to
follow a power law $\kappa\;S^{-\alpha}$ with
$\alpha=1.31^{+0.06}_{-0.03}$ (errors are the 10$^{\rm th}$ and
90$^{\rm th}$ percentiles) down to fluxes of $2\times 10^{-12}$ erg
cm$^{-2}$ s$^{-1}$, i.e. considerably below the BCS flux
limit. Although our best-fitting slope disagrees formally with the
canonical value of $-1.5$ for a Euclidean distribution, the BCS $\log
N-\log S$ distribution is consistent with a non-evolving cluster
population if cosmological effects are taken into account.

Our sample will allow us to examine large-scale structure in the
northern hemisphere, determine the spatial cluster-cluster correlation
function, investigate correlations between the X-ray and optical
properties of the clusters, establish the X-ray luminosity function
for galaxy clusters, and discuss the implications of the results for
cluster evolution.

\end{abstract}


\section{Introduction} 

Clusters of galaxies are the most massive entities to have collapsed
after decoupling from the Hubble expansion of the Universe. As a
consequence, they represent excellent tracers of the formation and
evolution of structure on the scale of a few to tens of Megaparsecs.

As far as the compilation of statistical samples of galaxy clusters is
concerned, the emphasis has, historically, been on optical cluster
properties such as the projected galaxy surface density in the cluster
core or the total number of galaxies in a pre-defined magnitude range
and within a fixed radius from the nominal cluster centre. The largest
compilations of this kind are the catalogues of Abell (1958), Abell,
Corwin \& Olowin (1989, from here on ACO), and Zwicky and co-workers
(1961--1968), each containing thousands of entries.

The statistical quality of optically selected cluster samples from
these early years is, however, seriously impaired not only by
projection effects but also by the subjectivity of an `eyeball'
selection process (Lucey 1983, Sutherland 1988, Struble \& Rood 1991).
Whereas this `human factor' has recently been eliminated almost
entirely by entrusting the detection and classification of clusters to
computer algorithms (e.g.\ Dalton et al.\ 1992, Lumsden et al.\ 1992,
Irwin, Maddox \& McMahon 1994, Wallin et al.\ 1994, Postman et al.\
1996), the fundamental problem remains that optical images probe the
distribution of galaxies only in two dimensions. Therefore,
fluctuations in the surface density of field galaxies as well as
superpositions of poor clusters along the line of sight can lead to an
overestimation of a system's richness. On the other hand, poor
clusters can be missed completely as they often do not contrast
strongly with the background field (Frenk et al.\ 1990, van Haarlem,
Frenk \& White 1997).  Extensive spectroscopic follow-up is required
to allow these effects to be quantified (e.g. Collins et al.\
1995). Future optical cluster surveys may be able to reduce the impact
of projection effects by using multi-colour photometry to improve the
contrast of the cluster galaxies over the foreground and background
galaxy population --- at present such techniques are still under
development.

The problem of projection effects can, however, be overcome almost
completely by selecting clusters in the X-ray rather than in the
optical band. X-ray emission from a diffuse, gaseous intra-cluster
medium (ICM) gravitationally confined at temperatures of typically a
few $10^7$ K is a sure indication that the system is indeed
three-dimensionally bound. Being due to ion-electron interactions and
thus proportional to the square of the density in the ICM, the X-ray
flux is also much more peaked at the gravitational centre of the
cluster than the projected galaxy distribution. This property requires
clusters to be almost perfectly aligned along the line of sight in
order to be mistaken for a single, more luminous entity, so that
projection effects caused by such superpositions can be effectively
neglected in the X-ray band.

X-rays thus provide a very efficient and less biased way of compiling
cluster samples. Early statistical cluster samples have been compiled
from the X-ray data taken by the {\sc Uhuru} (Schwartz 1978), {\sc
Ariel V} (McHardy 1978), and {\sc Heao 1 A-2} satellites (Piccinotti
et al.\ 1982), all of which comprised about 30 clusters. Including
clusters detected during the {\sc Exosat} mission, Edge et al.\ (1990)
presented an X-ray flux limited sample of 55 clusters, very similar in
size to the statistically complete $z\ge 0.14$ subset of the sample
compiled by Gioia et al.\ (1990) from X-ray sources detected in the
{\sc Einstein} Medium Sensitivity Survey (EMSS). The sample of Edge et
al.\ covers the whole extragalactic sky; its depth, however, is
limited to redshifts $z < 0.2$ due to the rather high flux limit at
$1.7 \times 10^{-11}$ erg cm$^{-2}$ s$^{-1}$ in the $2-10$ keV
band. Hence, only the more X-ray luminous systems are included. The
EMSS sample, on the other hand, extends to fluxes of $2 \times
10^{-13}$ erg cm$^{-2}$ s$^{-1}$ in the $0.3 - 3.5$ keV band and out
to redshifts of $z\sim 0.8$ but, even at the brightest fluxes, covers
no more than 740 square degrees, i.e., less than two per cent of the
sky. As a consequence, it samples preferentially an intermediate
luminosity range.

In 1991, the completion of the ROSAT All-Sky Survey (RASS, Voges 1992)
provided a database of unprecedented quality for the compilation of
X-ray selected cluster samples. Statistical cluster samples compiled
from RASS data over sky areas of the order of 1000 square degrees were
used by Nichol et al.\ (1994) and Romer et al.\ (1994) to compute the
cluster two-point correlation function.  On a larger scale, the first
RASS X-ray flux limited cluster sample with broad sky coverage has
recently been presented by Ebeling et al.\ (1996, hereafter
EVB). However, their all-sky sample (named XBACs) comprises only
ACO clusters, and, albeit X-ray {\em flux limited}, it is thus not
strictly X-ray {\em selected}; this applies also to the much smaller
sample of Nichol et al.\ (1994).

The ROSAT Brightest Cluster Sample (BCS) presented in this paper is
designed to overcome these limitations. The BCS is an X-ray flux
limited cluster sample covering the whole extragalactic sky
($|b|\geq 20^{\circ}$) in the northern hemisphere (where the median
exposure time in the ROSAT All-Sky Survey is 30 per cent longer than
in the South). At the same time, the BCS goes beyond optically known
clusters to include systems that were selected on the grounds of their
X-ray properties alone. We search explicitly for X-ray emission around
Abell and Zwicky clusters but also examine all bright X-ray sources
that were found to be significantly extended in the RASS. Optical
follow-up observations were performed, firstly, to confirm the cluster
nature of the X-ray selected systems, secondly, to obtain redshifts for
those BCS cluster for which reliable redshift measurements could not
be found in the literature, and, thirdly, to investigate the spectral
properties of the central cluster galaxies (Allen et al.\ 1992,
Crawford et al.\ 1995, Crawford et al.\ in preparation).

As the procedure followed in the compilation of the BCS is in many
respects very similar, if not identical, to the one employed during
the compilation of the XBACs sample, we shall limit the description of
certain steps to the minimum required to keep this paper
self-contained and, whenever possible, refer to EVB for a more
detailed account.

In this, logically if not chronologically, first article of a series
we present the BCS and discuss its statistical properties; in other
papers in this series we determine the X-ray cluster luminosity
function for the BCS and discuss cluster evolution (Ebeling et al.\
1997), constrain cosmological parameters using the BCS (Ebeling et
al., in preparation), and establish the spatial cluster-cluster correlation
function (Edge et al.\ in preparation).

Throughout this paper, we assume an Einstein-de Sitter Universe with
$q_0 = 0.5$ and $H_0 = 50$ km s$^{-1}$ Mpc$^{-1}$.

\section{The X-ray database$^{\dagger}$}
\footnotetext{\hspace*{-1mm}$^{\dagger}$ This short section is 
essentially the same as the identically named one in EVB; we feel 
that the gain in comprehensiveness of this paper justifies the 
repetition.}
\label{rass_db}

From 1990 August to 1991 February, the ROSAT X-ray satellite conducted
an all-sky survey (RASS) in the soft X-ray energy band ranging from
0.11 to 2.4 keV. Overviews of the ROSAT mission in general and the
RASS in particular can be found in the literature (Tr\"umper 1990,
Voges 1992).  A first processing of the data taken by the Position
Sensitive Proportional Counter (PSPC) during the RASS was performed as
the survey proceeded using the Standard Analysis Software System
(SASS, Voges et al.\ 1992) developed for this purpose at the
Max-Planck-Institut f\"ur Extraterrestrische Physik (MPE). To this
end, the incoming X-ray data were sorted into two-degree wide strips
of constant ecliptic longitude following the satellite's scanning
motion on the sky. Only three days worth of data are collected in one
strip, though, so that the exposure time in each strip is roughly
constant at some 360 s, and the much longer exposure times accumulated
around the ecliptic poles are not taken advantage of.

\begin{figure}
  \epsfxsize=0.5\textwidth
  \hspace*{0cm} \centerline{\epsffile{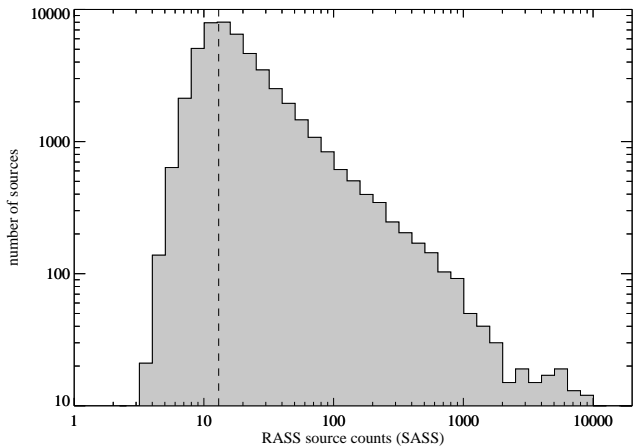}}
  \caption[]{The differential frequency distribution of the X-ray photon
	   counts in RASS sources as detected by the
	   SASS. Accordingly, the SASS master source list starts to
	   become incomplete at source strengths of less than about 15
	   photons. [Reproduced from EVB]} \label{sass_srccnts}
\end{figure}

Running on the ninety strips representing the RASS in this framework,
the SASS detected 49,441 X-ray sources (multiple detections removed,
see Cruddace et al.\ [1991] for details of the source detection
procedure) which constitute the X-ray sample upon which most of the
statistical RASS studies performed to date are based. Note that,
although, in this first processing of the survey data, the SASS
provided count rates in various energy bands, the actual source {\em
detection}\/ was performed exclusively in the PSPC broad band, i.e.,
in the energy range from $0.1-2.4$ keV. The source count distribution
for all SASS sources from this master source list\footnote{This is
exactly the same source list as the one used by EVB in the compilation
of the XBACs sample; Fig.~\protect\ref{sass_srccnts} is reproduced
from their paper.} is depicted in Fig.~\ref{sass_srccnts} and shows
that a minimum of some 10 to 15 photons is required for any source in
order to be detected reliably in this first processing of the RASS
data.  In terms of count rates, this translates into an approximate
detection limit of 0.04 to 0.05 count s$^{-1}$; by comparison, the
brightest RASS sources feature SASS count rates of the order of 10
count s$^{-1}$ (Voges 1992).

Since 1992 so-called PET files (for Photon Event Table) are produced at 
MPE, which contain the full photon information in a field of specified size 
around any given position in the sky thereby overcoming the limitations of the 
strip data used in the earlier processing.

Clearly, the optimal approach for the compilation of a highly complete
cluster sample from RASS data would be to run an algorithm optimized
for the detection of extended emission (e.g. VTP, see
Section~\ref{vtp_sample}) on PET fields for the whole study area.
However, running VTP on the whole northern extragalactic sky would
have required a total of almost 4,000 $2\times 2$ deg$^2$ PET fields
-- the production of such a large number of merged photon sets was not
feasible when this project was started. We therefore pursued the
approach described in detail in the following sections.

\section{A tentative sample}
\label{ten_sample}

\subsection{The SASS based sample}
\label{sass_sample}

\begin{figure}
  \epsfxsize=0.5\textwidth
  \hspace*{0cm} \centerline{\epsffile{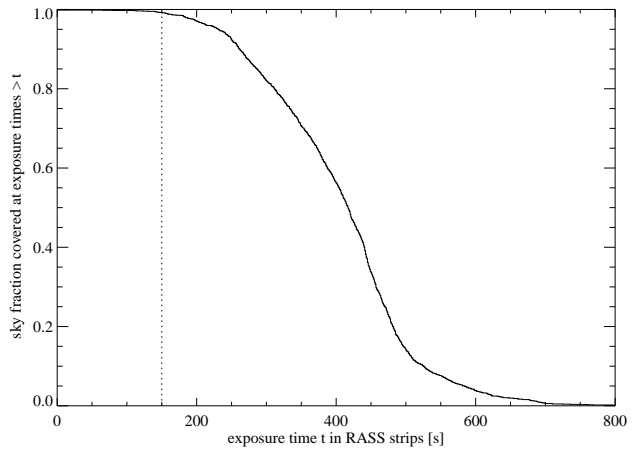}}
  \caption[]{The cumulative frequency distribution of the RASS exposure
	   times in the strips the first SASS processing was performed
	   on. Exposure times are computed in squares of 23 arcmin$^2$
	   area around 5000 positions randomly distributed on the
	   northern extragalactic sky.}  \label{rass_skycov}
\end{figure}

Starting from the results of the first SASS processing mentioned
above, we select a subset of 10,241 sources with SASS count
rates\footnote{For clusters of galaxies observed through a column
density $n_{\rm H}$ of neutral Hydrogen a PSPC broad band count rate
of 0.1 count s$^{-1}$ translates into an unabsorbed energy flux of
$1.2\times 10^{-12}$ erg cm$^{-2}$ s$^{-1} \times n_{\rm H}^{0.27}$
in the same $0.1-2.4$ keV band if $n_{\rm H}$ is given in units of
$10^{20}$ cm$^{-2}$. This formula is accurate to better than 10(5) per
cent for all gas temperatures in excess of 1.3(2.2) keV.} in excess of
0.1 count s$^{-1}$. This threshold is the result of a compromise
between maximal sky coverage and maximal survey depth: for sources
consisting of less than $\sim 15$ photons, the SASS source list starts
to become incomplete (see Fig.~\ref{sass_srccnts}), and at minimal
exposure times higher than 150 seconds the sky coverage of the RASS in
the northern hemisphere starts to fall below 99 per cent. The latter
figure is derived from the distribution shown in
Fig.~\ref{rass_skycov}. In the southern hemisphere the completeness
with respect to sky coverage is less than 89 per cent at a limiting
exposure time of 150 s (EVB) due to the fact that, in order to avoid
damage to the detector electronics, the PSPC was switched off
automatically during the all-sky survey when passing through regions
of enhanced particle background around the so-called South-Atlantic
Anomaly.

\begin{figure}
  \epsfxsize=0.5\textwidth \hspace*{0cm}
  \centerline{\epsffile{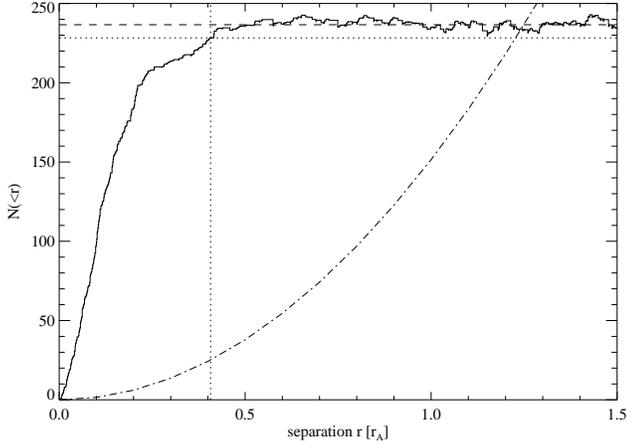}} 
  \caption[]{The cumulative number of coincidences in the
	    cross-correlation between the Abell catalogue and the RASS
	    X-ray source list provided by the SASS as a function of
	    the X-ray to optical source separation. Only X-ray sources
	    with SASS count rates higher than 0.1 count s$^{-1}$,
	    $\delta \geq 0^{\circ}$, and $|b| \geq 20^{\circ}$ are
	    considered. A parabolic background component (the
	    dot-dashed line) has been subtracted. The dashed line
	    marks the total number of true coincidences. The dotted
	    lines, finally, mark the selected maximal separation and
	    the corresponding number of true coincidences (see text
	    for details).}  \label{aco_sass_cumnum}
\end{figure}

We cross-correlated the reduced SASS X-ray source list with the Abell
cluster catalogue (limited to the northern extragalactic sky, i.e.\
$\delta \geq 0^{\circ}$, $|b| \geq 20^{\circ}$) using the clusters'
redshifts to scale angular to metric separations\footnote{Note that
both the angular and the metric separations discussed in this Section
are necessarily measured {\em in projection}\/ only.}. The latter are
measured in units of Abell radii ($1 r_{\rm A} = 1.5\,h_{\rm
100}^{-1}$ Mpc). Estimated redshifts based on the magnitude of the
tenth-ranked cluster galaxy, $m_{10}$, and taking into account the
cluster richness are used where measured redshifts are not available;
details of the procedure employed to estimate ACO cluster redshifts
can be found in EVB.  Fig.~\ref{aco_sass_cumnum} shows the resulting
cumulative distribution of the separations between the SASS X-ray
sources' and the optical cluster positions. Without any corrections
this distribution would be dominated by chance alignments whose number
increases as the area inside a radius $r$ and thus rises as the square
of the separation between the cluster and the respective X-ray source.
To counteract this effect, a parabolic contribution from random
coincidences fitted to the data in the $0.75 \leq r/r_{\rm A} \leq
1.5$ range has been subtracted (the dash-dotted curve in
Fig.~\ref{aco_sass_cumnum}).  The total number of true coincidences
(as opposed to chance coincidences) of 236.6 is given by the number at
which the cumulative distribution of Fig.~\ref{aco_sass_cumnum} levels
off and is indicated by the dashed line in Fig.~\ref{aco_sass_cumnum}.

Using the same cutoff in the respective separations as Ebeling and
co-workers for their XBACs sample ($r_{\rm max} = 0.407 \, r_{\rm A}$,
see the dotted lines in Fig.~\ref{aco_sass_cumnum}), we extract a 96.5
per cent complete sample consisting of 253 correlation pairs involving
252 unique SASS X-ray sources assigned to 235 different Abell
clusters. Here, like in the following, the quoted completeness (96.5
per cent) refers to the fraction of the total number of true
coincidences (marked by the dashed line in Fig.~\ref{aco_sass_cumnum})
found within the respective maximal radial separation.  As, even at
$z=0.4$, the mentioned radial cutoff of $r_{\rm max} = 0.407 \, r_{\rm
A}$ still translates into an angular separation of more than three
arcminutes, it always covers the uncertainty in the optical cluster
positions of typically 2 to 3 arcmin (ACO) which therefore do not need
to be taken into account separately. Some 25 entries from this list,
i.e.\ 10 per cent, can be expected to be coincidental.

Of the 49 catalogued Abell clusters in our study area with $z \leq
0.05$ only 30 are contained in this list. The lack of SASS detections
in excess of 0.1 count s$^{-1}$ for the remaining 19 nearby Abell
clusters is, however, not necessarily indicative of their being
intrinsically X-ray faint. Rather, it is possible that the SASS
detection algorithms have heavily underestimated the flux from such
potentially highly extended sources or may have missed the source
altogether (EVB), as the SASS was designed solely for the detection
of point sources.  We therefore also include the remaining 19 Abell
clusters with $z \leq 0.05$ which brings the total number of Abell
clusters in our tentative sample to 254.

As was shown by EVB, problems with the SASS detection efficiency as
far as cluster emission is concerned exist also for systems at
redshifts greater than 0.05. However, the sheer number of Abell
clusters at higher redshifts makes it unfeasible to include them all
as potential detections at this point (see Section~\ref{vtp_sample}
for a description of the re-processing of the RASS data performed at
the positions of all tentative SASS cluster detections). The amount of
incompleteness of the BCS introduced by missing SASS detections of
clusters at all redshifts is discussed in detail in
Section~\ref{serdet}.

\begin{figure}
  \epsfxsize=0.5\textwidth \hspace*{0cm}
  \centerline{\epsffile{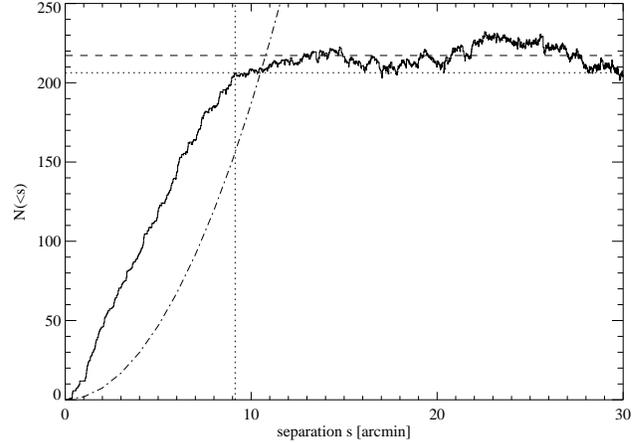}} 
  \caption[]{The cumulative number of coincidences in the
	    cross-correlation between the Zwicky catalogue and the
	    RASS X-ray source list provided by the SASS as a function
	    of the X-ray to optical source separation. Only X-ray
	    sources with SASS count rates higher than 0.1 count
	    s$^{-1}$, $\delta \geq 0^{\circ}$, and $|b| \geq
	    20^{\circ}$ are considered. A parabolic background
	    component (the dot-dashed line) has been subtracted.  The
	    dashed line marks the total number of true
	    coincidences. The dotted lines, finally, mark the selected
	    maximal separation and the corresponding number of true
	    coincidences (see text for details).}
	    \label{zwi_sass_cumnum}
\end{figure}

For the 8528 Zwicky cluster with $\delta \geq 0^{\circ}$ and $|b| \geq
20^{\circ}$ the procedure is essentially the same. Note, however, that
we now perform the cross-correlation with the SASS X-ray source list
on an angular rather than a metric scale, as redshifts (measured or
estimated) are not available for the whole of the Zwicky
catalogue. Just as Fig.~\ref{aco_sass_cumnum} does for Abell clusters,
Fig.~\ref{zwi_sass_cumnum} summarizes the results of the
cross-correlation between the 10,241 RASS X-ray sources with SASS
count rates above 0.1 count s$^{-1}$ and the Zwicky cluster list.  The
total number of true coincidences in this cross-correlation is 217.2;
a 95 per cent complete sample extracted at separations of less than
9.1 arcmin consists of 362 coincidences (involving 355 unique SASS
sources and 354 different Zwicky clusters) of which some 156 (43 per
cent) are expected to be non-physical.

\begin{figure}
  \epsfxsize=0.5\textwidth \hspace*{0cm}
  \centerline{\epsffile{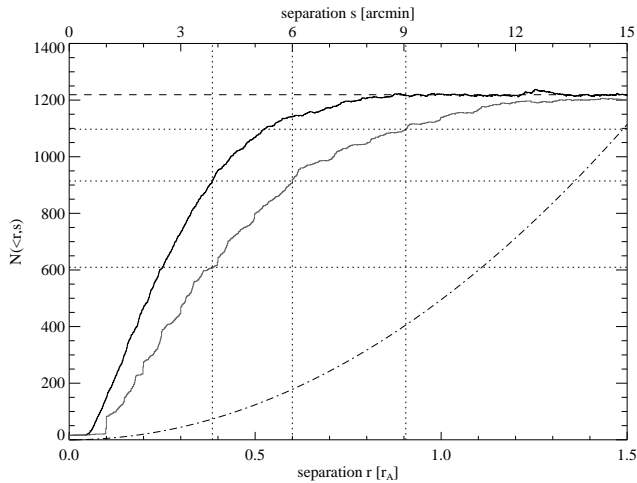}} 
  \caption[]{The cumulative number of coincidences in a
  	    cross-correlation between the Zwicky and the Abell
  	    clusters in our study area as a function of their
  	    separation. The bold solid curve represents the number of
  	    coincidences within a given metric separation, the fine
  	    line corresponds to angular separations as annotated on
  	    the upper x axis. A parabolic background component (the
  	    dot-dashed line) has been subtracted. The dashed line
  	    marks the total number of true coincidences. The dotted
  	    lines, finally, mark the 50$^{\rm th}$, 75$^{\rm th}$, and
  	    90$^{\rm th}$ completeness percentiles for the set of true
  	    coincidences found in the angular cross-correlation.}
  	    \label{zwi_aco_cumnum}
\end{figure}

Because of the substantial overlap between the Abell and Zwicky
cluster catalogues, many of the X-ray sources associated with Zwicky
clusters in our cross-correlation are, at the same time, also
contained in the subsample extracted from the cross-correlation with
the Abell catalogue.  Fig.~\ref{zwi_aco_cumnum} shows the cumulative
number of coincidences in a positional cross-correlation between the
Abell and Zwicky clusters in our study area as a function of both
their metric (thick solid line) and angular separations (thin solid
line)\footnote{The conversion from angular to metric separations is
performed at the redshift of the respective Abell cluster}. According to
Fig.~\ref{zwi_aco_cumnum}, some 1,200 clusters have been detected
independently by Abell and Zwicky and co-workers.  Although, for half
of them, the optical cluster centroids determined by Abell and Zwicky
agree to within about four arcminutes, one quarter of the common
clusters are separated by between four and six arcminutes, and for 10
per cent the nominal cluster positions differ by more than nine
arcminutes or more than 0.5 Abell radii.

In order to eliminate multiple entries from our tentative sample of
RASS detected Abell and Zwicky clusters, we exclude from our list all
Zwicky clusters that lie within 10 arcmin of an Abell cluster, {\em if
both have been assigned the same X-ray source in the
cross-correlations}. Having thus eliminated 137 of the 354 Zwicky
clusters in our list, we are left with a tentative sample of 446 Abell
and Zwicky clusters with SASS count rates of at least 0.1 count
s$^{-1}$.  Including the 19 nearby Abell clusters that might have been
missed by the SASS we arrive at a total of 465 candidate clusters for
the BCS.

So far, all our candidate clusters have been selected from optical
cluster catalogues which can be trusted to be reasonably complete for
optically rich clusters at redshifts $z \la 0.2$ (Huchra et al.\ 1990,
Scaramella et al.\ 1991). However, at redshifts higher than about 0.2,
even very rich clusters start to become inconspicuous on the optical
plates used in the compilation of these catalogues. Incompleteness is
a potential problem also at the low redshift end ($z\la 0.07$) where
any X-ray flux limited sample (such as the BCS) will be dominated by
low-luminosity systems that may not be optically rich enough to be
included in optical catalogues. This effect of a thinning-out at low
redshifts can be clearly seen in the $L_{\rm X}-z$ distribution of the
XBACs (EVB).

\begin{figure}
  \epsfxsize=0.5\textwidth  \hspace*{0cm}
  \centerline{\epsffile{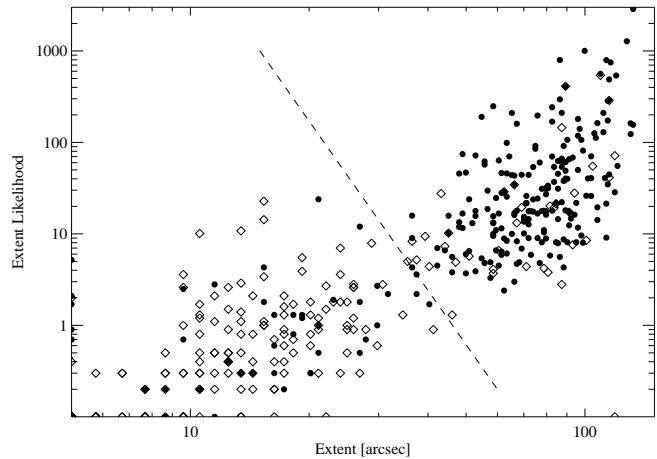}}
  \caption[]{The distribution of the SASS source parameters X--ray
	   extent likelihood and X-ray extent for rich Abell clusters
	   (filled circles) and random field sources (open diamonds).
	   Note that, while both samples comprise the same number of
	   sources, many of the randomly selected sources have X-ray
	   extents of zero and do not feature on this logarithmic
	   plot.  [Reproduced from Ebeling et al.\ (1993)]}
	   \label{exexl}
\end{figure}

To overcome these problems, we use the X-ray source extent to find
galaxy clusters not included in the Abell and Zwicky catalogues. The
extent parameter supplied by the SASS is the Gaussian width of the
radial source profile that is in excess of the instrumental
resolution, i.e.\ the PSPC point spread function (PSF) for the all-sky
survey. Although a Gaussian is not really an adequate description of
the surface brightness distribution of extended X-ray sources such as
clusters of galaxies, the SASS extent parameter is nonetheless very
useful for flagging possibly extended emission. As was shown by
Ebeling et al.\ (1993), RASS X-ray sources with a SASS extent value in
excess of 35 arcsec are almost exclusively clusters of galaxies
(Fig.~\ref{exexl}).  We emphasize that while, at high Galactic
latitude, a significant SASS extent comes close to being a sufficient
criterion for a cluster identification (it is not perfect due to the
possibility of blends of close source pairs), it is not a necessary
condition.  The latter is demonstrated by a number of Abell clusters
classified as point sources by the SASS (see the lower left of
Fig.~\ref{exexl}) and is discussed in depth in
Section~\ref{sassvsvtp}.

We therefore select from the list of SASS detections with count rates
greater than 0.1 count s$^{-1}$ all 406 sources that fall within our
study area ($\delta \geq 0^{\circ}$, $|b| \geq 20^{\circ}$) and
feature a SASS extent of at least 35 arcsec. 218 of these are already
included in our list on the grounds of their being associated with
Abell or Zwicky clusters; another eleven can be readily discarded as
multiple detections of very bright, nearby sources like M\,87. We add
the remaining 177 to our list to finally obtain a sample of 642 BCS
candidates.

For all of these we request from the RASS data archive PET data in
$2\times 2$ deg$^2$ fields around the optical cluster position or the
position of the extended SASS source. As the PET file for any field
contains all photons accumulated during the whole RASS in the
respective region, the accumulated exposure times in these fields are
sometimes much longer than those available in the SASS strips; up to
8200 s are attained. The median value of 460 s, however, is very
similar to the one for the initial strips.

\subsection{The VTP based sample}
\label{vtp_sample}

A fresh search for X-ray emission from clusters of galaxies in these
data is now required not only because of the greater depth of the PET
maps; we also want to overcome the limitations of the SASS detection
algorithms that have been used on the RASS strip data. As mentioned
before, the SASS has difficulties characterizing or, for that matter,
even reliably detecting extended emission. The cause for this lies in
the algorithm's basic design as a point source detection algorithm
which is primarily sensitive to highly localized and quasi-spherical
intensity variations. We therefore re-process the RASS data in the PET
fields with VTP (for Voronoi Tesselation \& Percolation), an algorithm
developed for the detection and characterization of sources of
essentially arbitrary shape (Ebeling \& Wiedenmann 1993, Ebeling
1993). VTP works on the individual photons rather than on spatially
binned representations of the data; also, it does not require any
model profiles to be fitted to the detected emission in order to
determine fluxes, all of which makes the algorithm extremely versatile
and flexible. The re-processing of the data with VTP is crucial in
order to obtain reliable fluxes for the clusters in our sample: for
clusters of galaxies, raw SASS fluxes are found to be typically too
low by about a factor of 0.5; for nearby clusters the misassessment
can even exceed one order of magnitude (EVB).

At the time of writing, more than 98 per cent of our 642 BCS
candidate clusters have had their PET fields processed by VTP. For 11
sources, PET files are either not available or their VTP processing
has been delayed due to very high photon counts in a few fields at
high ecliptic latitude which cannot be processed in one go by VTP. In
order to assess the impact of missing fields on the completeness of
the BCS, we have tried to identify the corresponding SASS sources. We
find that
\begin{itemize}
  \item The only source of these eleven that is associated with an Abell
        cluster (A\,2506) is 0.37 $r_{\rm A}$ away from the optical
        cluster centre and coincides within 7 arcseconds with a close
        pair of bright stars of $m_{\rm V} \sim 7$.
  \item Of the eight sources tentatively assigned to Zwicky clusters,
	six feature X-ray/optical separations of 9 arcmin or more and
	are thus very likely to be chance coincidences with
	non-cluster sources. Indeed, three of them have firm stellar
	identifications, another two coincide with catalogued AGN, and
	only for one the identification (although clearly non-cluster)
	is ambiguous in as much as the source is probably a blend of
	emission from an IRAS galaxy and NGC\,6521.
  \item The two sources without VTP results from our list of 177 sources
        classified as extended by the SASS are identified with a star
        and a QSO, respectively. With SASS count rates in excess of
        0.3 count s$^{-1}$ both of these point sources are
        sufficiently bright for the instrumental PSF to be sampled
        with good statistical accuracy. The simple approximation to
        the PSF used by the SASS (a single Gaussian) is known to lead
        to spurious extents for bright point sources. Also, both
        sources lie within two degrees of the north ecliptic pole
        (NEP), so that small inaccuracies in the attitude solution of
        the ROSAT satellite in the immediate vicinity of the NEP may
        contribute to the erroneous extent classification.
\end{itemize}
Thus, nine of the eleven sources for which VTP results are not
available are associated with non-cluster counterparts. The two
candidates left that might be true cluster detections missing from our
tentative list are both Zwicky clusters, namely Z\,3803 and
Z\,8417. The former is a previously detected EMSS cluster, and the
latter, although previously undetected, has a galaxy distribution
consistent with a cluster at a redshift $z\ga 0.2$. Thus both are
accepted as cluster identifications. Summarizing we can say that the
incompleteness of two per cent in our BCS/VTP database has no
significant implications for the completeness of the BCS as a whole.

The sources detected in the 631 PET fields processed by VTP are then
screened for possible blends of close source pairs. Such blends occur
because VTP makes no assumptions about the intrinsic surface
brightness distribution of any source that it considers to be a
significant enhancement over the background. As a consequence, the VTP
percolation algorithm needs to be forced to stop at the interface of
two sources whose detected surface brightness profiles overlap, in
order to keep the two sources separated. Deblending was required for
55 VTP sources. Note that direct deblending is only feasible for
bright blends whose components are clearly discernible as separate
sources on the RASS images (see Fig.~8 of EVB for an illustration of
the deblending procedure). The majority of blends, however, occur near
the detection limit and typically involve sources that are too faint
to be resolved in the RASS. Falling well below the flux limit, these
faint blends are of no immediate concern for the BCS proper; however,
if not recognized, they would contribute significantly to the number
of X-ray extended sources at the faint end of the cluster $\log
N$-$\log S$ distribution presented in Section~\ref{logNlogS}.  Our
procedure of identifying and eliminating unresolved blends is
discussed in Section~\ref{clean_up}.

Just as before with the SASS sample, we then introduce a count rate
cut for the VTP detections thereby eliminating sources at flux levels
at which any sample would be intrinsically incomplete due to the
limited depth of the RASS alone. A threshold value of 0.07 count
s$^{-1}$ is chosen for this purpose. As will be shown in
Section~\ref{logNlogS}, VTP sources below this count rate threshold
can be safely ignored for our purposes as, statistically, fewer than
one these faint sources would reach final energy fluxes in excess of
$2\times 10^{-12}$ erg cm$^2$ s$^{-1}$ ($0.1-2.4$ keV), which is the
flux level at which the RASS starts to become incomplete in the
northern hemisphere due to exposure time limitations. Application of
this count rate limit leaves a sample of 1650 VTP sources.

\begin{figure}
  \epsfxsize=0.5\textwidth
  \hspace*{0cm} \centerline{\epsffile{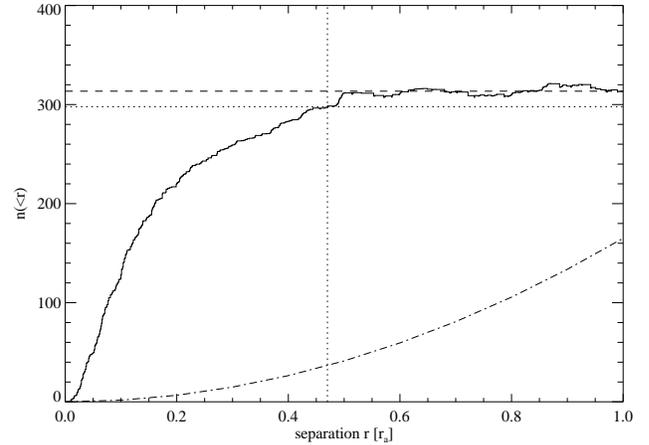}}
  \caption[]{The cumulative number of coincidences in the
	   cross-correlation between the ACO catalogue and the RASS
	   X-ray source list provided by VTP as a function of the
	   X-ray to optical source separation. Only X-ray sources with
	   VTP count rates higher than 0.07 count s$^{-1}$ are
	   considered.  A parabolic background component (the
	   dot-dashed line) has been subtracted. The dashed line marks
	   the total number of true coincidences. The dotted lines,
	   finally, mark the selected maximal separation and the
	   corresponding number of true coincidences.}
	   \label{aco_vtp_cumnum}
\end{figure}

Subsequently, the merged VTP X-ray source list for all PET fields is
correlated against the Abell catalogue, and, just as before, a 95 per
cent complete sample of coincidences is compiled (see
Fig.~\ref{aco_vtp_cumnum}). The maximal separation between the
included matches is 0.470 Abell radii. This sample now comprises 334
unique X-ray sources corresponding to 298 different Abell clusters
with some 11 per cent of the entries being caused by positional chance
coincidences (as indicated by the dot-dashed line in
Fig.~\ref{aco_vtp_cumnum}).

\begin{figure}
  \epsfxsize=0.5\textwidth
  \hspace*{0cm} \centerline{\epsffile{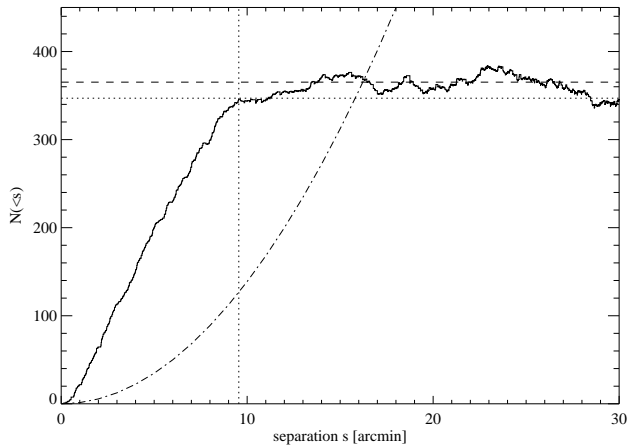}}
  \caption[]{The cumulative number of coincidences in the
	   cross-correlation between the Zwicky cluster catalogue and
	   the RASS X-ray source list provided by VTP as a function of
	   the angular distance between the X-ray source and the
	   optical cluster position. Only X-ray sources with VTP count
	   rates higher than 0.07 count s$^{-1}$ are considered.  A
	   parabolic background component (the dot-dashed line) has
	   been subtracted. The dashed line marks the total number of
	   true coincidences. The dotted lines, finally, mark the
	   selected maximal separation and the corresponding number of
	   true coincidences.}  \label{zwi_vtp_cumnum}
\end{figure}

For the Zwicky clusters the cross-correlation with the VTP source list
results in 365 true coincidences. Fig.~\ref{zwi_vtp_cumnum} shows
the cumulative number of coincidences as a function of angular
separation. A 95 per cent complete sample consists of 461 unique X-ray
sources associated with 461 Zwicky clusters at separations of up to
9.55 arcmin. Just as before for the tentative sample from the
cross-correlation with the SASS source list, the contamination of this
sample by chance coincidences is high: almost 27 per cent of the
correlation pairs within a maximal separation of 9.55 arcmin are
expected to be coincidental.

Note that the number of true coincidences found in the
cross-correlations (the number at which the cumulative distributions
of Figs.~\ref{aco_sass_cumnum}, \ref{zwi_sass_cumnum},
\ref{aco_vtp_cumnum} and \ref{zwi_vtp_cumnum} level off) is
considerably higher when the VTP source list is used. More than 60 per
cent more Zwicky clusters are found in the cross-correlation with the
VTP source list than in the previous one with the original SASS list;
for Abell clusters the increase in detections is less dramatic but
amounts still to more than 30 per cent. The reasons for this rise are
twofold: firstly, and most importantly, the VTP source list has a
lower count rate limit than the SASS list. Consequently, a number of
(mostly faint) X-ray sources get included in the VTP based sample as
serendipitous detections of Abell and Zwicky clusters in our PET
fields.  This effect is amplified by the presence of a few PET fields
with exposure times much longer than those of the strip data used in
the first SASS analysis. A second cause for the increase in cluster
detections lies in the source detection procedures themselves as
nearby low-surface brightness sources missed or misassessed by the
SASS are now detected by VTP.

Again, there is considerable overlap between the X-ray source lists
obtained from the correlations between the VTP source list and the
Abell and Zwicky cluster catalogues, respectively. Merging the two
samples yields a list of 612 unique VTP sources with count rates of at
least 0.07 count s$^{-1}$ that have been associated with Abell or
Zwicky clusters in the extragalactic part of the northern hemisphere.

\begin{figure}
  \epsfxsize=0.5\textwidth
  \hspace*{0cm} \centerline{\epsffile{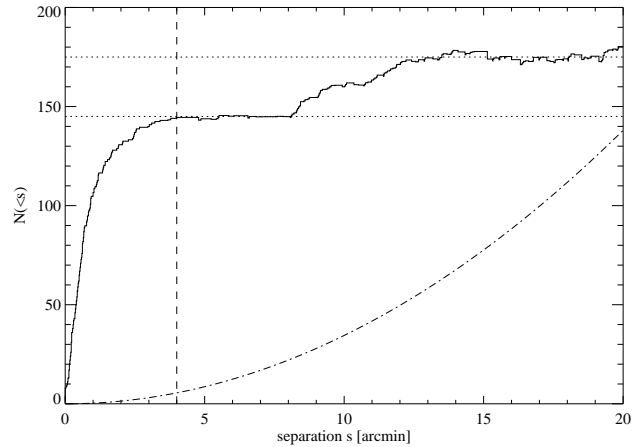}}
  \caption[]{The cumulative number of coincidences in the
	   cross-correlation between the SASS extended sources that do
	   not coincide with either Abell or Zwicky clusters and the
	   RASS X-ray source list provided by VTP as a function of the
	   angular distance between the SASS and VTP X-ray source
	   position. A parabolic background component (the dot-dashed
	   line) has been subtracted. Note the jump in the total
	   number of true coincidences highlighted by the dotted
	   lines. The dashed vertical line marks the maximal allowed
	   separation for sources accepted as real detections of
	   potentially extended X-ray sources.}
	   \label{sxtd_vtp_cumnum}
\end{figure}

\label{sass_crash}
We still have to incorporate the list of sources flagged as extended
by the SASS. As the PET fields around them have also been re-processed
by VTP, we now have to find the VTP counterparts of the SASS sources
the PET fields are centred upon. This leads to the inclusion of
another 132 VTP sources with count rates in excess of 0.07 count
s$^{-1}$ that lie within four arcmin of the original SASS position and
have not yet been included as possible detections of Abell or Zwicky
clusters. Note that the number of VTP detections of SASS extended
sources thus included (132) is considerably lower than the number of
extended (non-Abell and non-Zwicky) SASS sources we started from (177,
for 175 of which we have PET fields). This discrepancy is mainly due
to some two dozen `SASS crashes', i.e., detections of, in general,
very bright point sources that the SASS erroneously flagged as
extended.  Since these sources also have large positional errors of
typically 10 arcmin, they are excluded from entering our sample by the
cutoff at a respective SASS-VTP source separation of four arcmin.
Fig.~\ref{sxtd_vtp_cumnum} illustrates the clear separation between
regular SASS detections and `SASS crashes'. Note the significant rise
in the number of true coincidences at some 10 arcmin caused by bright
point sources with large errors in the SASS source position.

Having included the VTP counterparts of bright X-ray sources
classified as extended by the SASS, we are left with one more class of
potential cluster detections, namely the sources classified as
extended by VTP that are not yet included. Since, contrary to the
SASS, VTP does not assume any model for the radial source profile in
the source detection procedure, it necessarily can not provide an
immediate measure of the source extent.  However, VTP does determine
source extents at a later stage when correcting the detected source
count rates for low-surface brightness emission that has escaped
direct detection. In order to ensure greatest possible completeness
for our tentative sample, we use a deliberately low threshold for the
VTP source extent of $r_c = 15$ arcsec (see Section~\ref{flux_corr}
below) and add another 439 X-ray sources to our list, bringing the
total up to 1183 VTP sources. Including in our list of cluster
candidates all sources classified as extended by either the SASS or
VTP also allows us to quantitatively assess the efficiency of
selecting a cluster sample purely by the sources' X-ray
characteristics (see Section~\ref{xraysel}).
  
\section{Count rate corrections and the determination of VTP 
         source extents}
\label{flux_corr}

In the presence of background radiation, the emission directly
detectable by any source detection algorithm is always limited to some
fraction of the total flux. Hence, the raw VTP count rates have to be
corrected for the low surface brightness emission in the far wings of
the source that has not been detected directly. While usually small
for point-like sources, the `missing' fraction of the total projected
source emission can be substantial for extended sources such as
clusters of galaxies. A correction to the directly detected emission
like the one outlined in the following plays thus a crucial r$\hat{\rm
o}$le in the compilation of any X-ray cluster sample.

Since the procedure employed to this end is described in detail by
EVB, we give only a brief summary here.  The algorithm uses two
different kinds of source profile: The first one applies to extended
cluster emission that is resolved in the RASS and assumes a King
model, i.e., a radial surface brightness profile of the form
\[
 \sigma(r) = \sigma_{\rm K}(r) = \sigma_0 \left[ 1+(r/r_c)^2 
                                 \right]^{-3\beta+1/2} 
\]
(Cavaliere \& Fusco-Femiano 1976) where $\sigma(r)$ is the projected
surface brightness as a function of radius. Fixing the beta parameter
at a value of 2/3 (Jones \& Forman 1984), we can derive both the
normalization $\sigma_0$ and the core radius $r_c$ from the VTP source
characteristics. Note that it is only in {\em correcting}\/ the
detected count rates that a specific model for the source profile and,
in particular, spherical symmetry is assumed for VTP detections.

The second source profile assumed in the count rate correction 
procedure is that of a point source, i.e.
\[
 \sigma(r) = \sigma_{\rm ps}(r) = \frac{s_{\rm ps}}{2\pi r}\, \delta(r)
\]
where $s_{\rm ps}$ is the total source count rate and $\delta(r)$ is
Dirac's delta function. This model is applied to cluster emission that
is not resolved as extended in the RASS (see below and
Section~\ref{extent} for a discussion of the VTP extent parameter).

The expected observed surface brightness distribution
$\tilde{\sigma}(r)$ is the convolution of the source's intrinsic
surface brightness distribution $\sigma(r)$ with the instrument
response:
\begin{equation}
\tilde{\sigma}(r) = \int_0^{2\pi} \int_0^\infty
                  \sigma(|\vec{r}-\vec{r}\,'|) 
                  \,{\rm PSF}(r') \, r' \, dr' \, d\phi.
\label{eq_sigma}
\end{equation}
For the RASS, the telescope's point-spread function, PSF($r$), is the
weighted average of the PSFs at all off-axis angles. PSF($r$) has been
computed for several photon energies by Hasinger et al.\ (1994); we
employ a numerical representation of PSF($r$) for $E=1$ keV (De
Grandi, private communication) that was shown to be in excellent
agreement with the source profiles found for RASS-detected AGN
(Molendi, private communication).

As described in more detail in EVB, the free parameters of the model
profile $\tilde{\sigma}(r)$ can be derived from observed source
characteristics without any radial fitting. The quantities used are
the total detected count rate which must equal the integral over the
model profile out to the effective detection radius, and the surface
brightness level relative to the background, used as the equivalent
of a percolation radius in VTP's percolation step.

The true total source count rate can then be determined from 
\begin{equation}                                                               
s_{\rm true} = 2 \pi \int_0^\infty \sigma_{\rm K}(r) \, r \, dr 
             = \frac{\pi \, \sigma_0 \, r_c^2}{3\,(\beta-1/2)}.
\label{king_true}
\end{equation}                            
(for clusters of galaxies) or
\begin{equation}                                                               
 \frac{s(r_{\rm VTP})}{s_{\rm true}} = 
           2 \pi \int_0^{r_{\rm VTP}} {\rm PSF}(r) \, r \, dr.
\label{ps_true}
\end{equation}                                                               
(for point-like sources).

Clearly, our assumption that the observed emission from any cluster
can be described by either of these model profiles represents an
oversimplification, in particular in the presence of cooling flows.
However, at no stage of the flux correction procedure is the model
profile from equation~\ref{eq_sigma} fitted to the actually observed
surface brightness distribution. Although we do assume a specific
model, it is only its integral properties that enter, which is why
deviations of the true distribution from the assumed model (such as
the extremely peaked surface brightness profile observed in the very
core of cooling flow clusters) have much less effect on the result
than they do for the fitting procedures employed by conventional
source detection algorithms. As, in all but the largest cooling flow
clusters, it is the emission from outside the cooling core rather than
from the cooling flow itself that dominates the cluster's overall
X-ray luminosity (White, Jones \& Forman 1997), we are confident that
our integral correction to the total emission will not be largely
affected by the presence of cooling flows (this is confirmed by
Fig.~\ref{RASS_PO_cr_corr} which is discussed later in this section).

However, as far as the initial SASS detections are concerned, the
presence of cooling flows may cause some bias. As mentioned before,
the SASS was designed as a point-source detection algorithm. Thus,
cooling-flow clusters are likely to be preferentially detected and
therefore over-represented in the sample of SASS clusters that our
analysis started from. We have no means to correct for this bias other
than to rely on VTP to serendipitously detect low surface brightness
emission from non-cooling-flow clusters that the SASS has missed.
Section~\ref{serdet} discusses the importance of these serendipitous
detections in detail. Once a cluster is detected (or re-detected) by
VTP, the presence of a cooling flow will not greatly affect the flux
correction procedure.

The lack of any radial fitting in the count rate correction procedure
entails that for any particular cluster the value for the cluster core
radius $r_c$ determined in the flux correction process cannot be
expected to be as accurate as the results of a detailed imaging
analysis of pointed data. Rather, it should be seen as a statistical
parameter that allows the extent of the source and the fraction of the
cluster emission that has escaped direct detection to be assessed.  In
the following, we will use the term `VTP extent' as a synonym for
$r_c$.

\begin{figure}
  \epsfxsize=0.5\textwidth
  \hspace*{0cm} \centerline{\epsffile{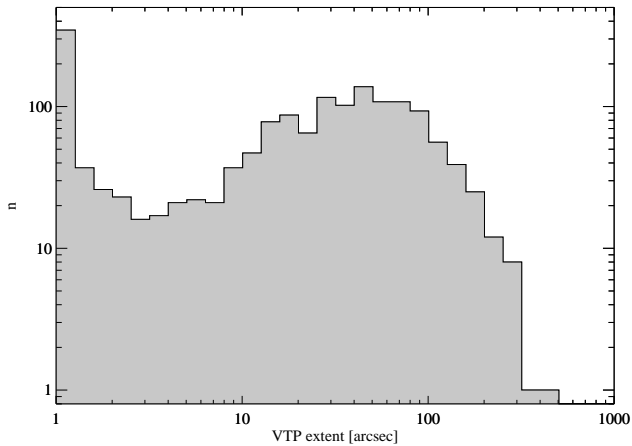}}
  \caption[]{The distribution of VTP source extents for the 1650 sources
	   in our PET fields with VTP detected count rates in excess
	   of 0.07 count s$^{-1}$. Sources with extent values below
	   the plotted range have been assigned an extent of 1
	   arcsec. Note the gap in the distribution indicating that
	   source extents of less than about 10 arcsec are not
	   resolved in the RASS.}  \label{bcs_vtp_extent}
\end{figure}

\begin{figure}
  \epsfxsize=0.5\textwidth
  \hspace*{0cm} \centerline{\epsffile{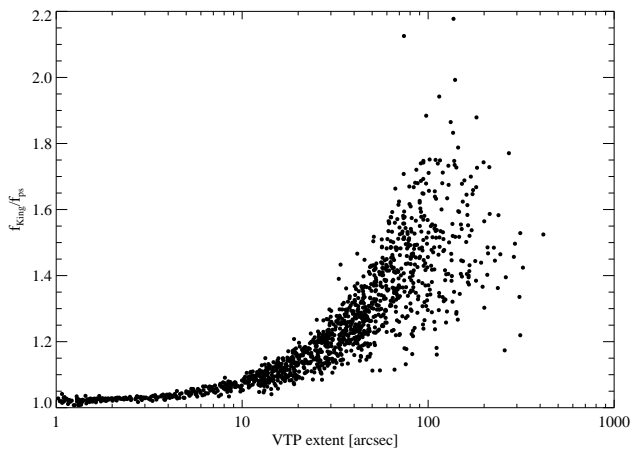}}
  \caption[]{The ratio of the count rate correction factors (King
  	     model/point source) for the sources of
  	     Fig.~\protect\ref{bcs_vtp_extent} as a function of VTP
  	     extent, i.e., core radius. Note how the correction
  	     factors for the two models converge for vanishing extent
  	     values.}  \label{fc_rc}
\end{figure}

\begin{figure}
  \epsfxsize=0.5\textwidth
  \hspace*{0cm} \centerline{\epsffile{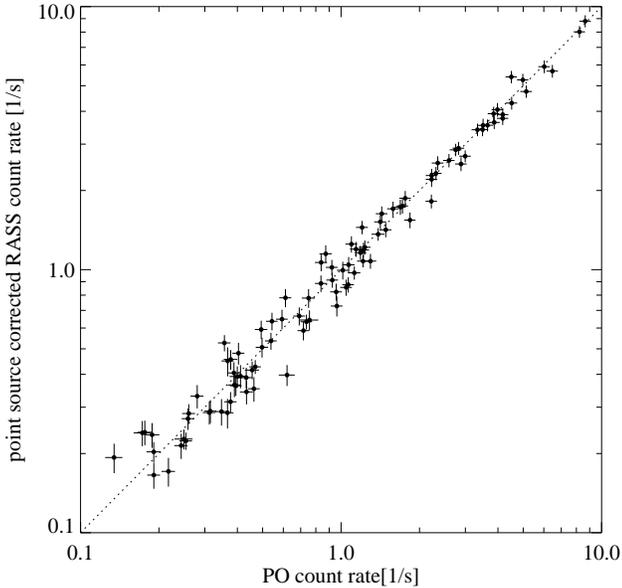}}
  \caption[]{PSPC broad band count rates for 100 ACO galaxy clusters as
	     derived from ROSAT All-Sky Survey data using VTP compared
	     to the corresponding values obtained from pointed
	     observations (PO). Note that the contribution from point
	     sources is {\em excluded}\/ in the PO data.  The error
	     bars allow for a five per cent systematic uncertainty in
	     addition to the Poissonian errors.  [Reproduced from
	     EVB]} \label{RASS_PO_cr_corr}
\end{figure}

The distribution of VTP extents ($r_c$) for the 1650 sources detected
by VTP at count rates in excess of 0.07 count s$^{-1}$ in the 631
processed PET fields is shown in Fig.~\ref{bcs_vtp_extent}. Note that
the distribution peaks at about 40 arcsec indicating that sources with
an extent parameter considerably below this value are not reliably
resolved as extended in the RASS. When applying the flux correction
factors, we take the conservative approach to consider only sources
with an extent of less than 15 arcsec to be point-like and apply
equation~\ref{ps_true}, whereas sources featuring VTP extents greater
than this threshold value are considered to be potential clusters
whose count rates are corrected according to
equation~\ref{king_true}. As the results from Eqs.~\ref{king_true} and
\ref{ps_true} converge as the source extent obtained from the King
model decreases (see Fig.~\ref{fc_rc}) and differ typically by only 10
per cent or less for extent values smaller than 15 arcsec, is does not
really matter where exactly the line between point sources and
extended sources is drawn. Note, however, that even for the most
extreme case of a very compact cluster at a redshift of $z=0.5$ (which
is higher than that of any cluster detected so far in the RASS) a
metric core radius of only 150 kpc would still correspond to more than
20 arcsec on an angular scale. Thus we do not believe that the use of
the correction factor from the convolution of the King model with the
RASS PSF is scientifically justifiable at extent values significantly
below $r_c\sim 15$ arcsec.

It ought to be noted here that we do not expect all sources with VTP
extents in excess of 15 arcsec to be really physically extended.
Undoubtedly, a considerable fraction of the sources with extents
between about 10 and 40 arcsec will be blends near the VTP detection
limit. These will be eliminated from our sample at a later stage when
we assess the physical identifications of all sources in our tentative
sample from their combined X-ray/optical appearance.  At this point we
are not yet concerned with source identification -- rather, we want to
mark {\em potentially}\/ extended sources for inclusion in our cluster
candidate list. At the same time, the above procedure provides us with
total count rates (fluxes) for all sources that will finally remain in
our sample.

While the count rate correction procedure outlined above recovers
emission that has escaped direct detection for point-like and extended
sources alike, it assumes a single surface brightness profile and does
thus not account for emission from more than one contributor. For
clusters of galaxies this means that any contribution from point
sources (usually AGN) or cooling flows to the overall emission will
result in an overestimation of the {\em detected}\/ count rate if
interpreted wholly as emission from an isothermal ICM. Subsequently,
the overestimation of the {\em corrected}\/ count rate will be even
greater as both the diffuse and the point-like emission get corrected
assuming the same King model. Hence, the raw (detected) VTP count
rates for clusters will be too low and the corrected ones will be too
high\footnote{Using VTP results obtained for data from simulated PSPC
cluster observations, Scharf et al.\ (1997) recently pointed out that
the tendency for the corrected VTP count rates to be too high may in
part also be attributed to `positive background noise', i.e., upward
fluctuations in the observed background photon density, around the
cluster that gets included in the VTP detection.}.  Since, at the
angular resolution and sensitivity of the RASS, it is impossible to
quantify the point source or cooling flow contribution to the X-ray
emission from all but the brightest and most nearby clusters of
galaxies, we resort to a final heuristic correction to account for
this problem by adopting the geometric mean of the detected and the
corrected count rates as the best VTP estimate of the true count rate
from diffuse ICM emission.

The accuracy of these final RASS-VTP count rates of galaxy clusters
can be assessed by comparing them to the PSPC broad band count rates
obtained in pointed observations (POs) of the same clusters. Such a
comparison has been made by EVB for their XBACs, the fluxes of which
were derived in exactly the same way as the ones for the BCS.
Fig.~\ref{RASS_PO_cr_corr} shows RASS-VTP and PO count rates for 100
ACO clusters for which both RASS and pointed data are available. Note
the excellent agreement between the VTP count rates based on RASS
exposure times of typically 400 s and the values obtained from pointed
PSPC observations that are typically 25 times deeper (median PO
exposure: 10.2 ks). Since the sample of 100 clusters used in this
comparison contains both clusters with and without cooling flows, the
good agreement between the RASS-VTP count rates and the total
(aperture) count rates obtained from pointed data confirms our earlier
statement that the presence of cooling flows does not greatly affect
our flux correction procedure. Using the data shown in
Fig.~\ref{RASS_PO_cr_corr} to calibrate the $1\sigma$ uncertainty
$\Delta s$ in the RASS-VTP count rates, EVB find
\begin{equation}
   \frac{\Delta s}{s} = 2.29 \, (s\,t_{\rm exp})^{-0.48},
   \label{vtp_err}
\end{equation}
where $s$ is the final VTP count rate and $t_{\rm exp}$ is the RASS
exposure time.

The overall agreement between VTP and PO fluxes is thus very good.
However, one might suspect that the $2\times 2$ deg$^2$ fields that
the VTP analysis was performed in could be too small to fully cover
the very extended ICM emission from the most nearby clusters of
galaxies, resulting in too low VTP fluxes.  Inspecting the RASS X-ray
images of bright and nearby clusters, we find that our limited field
size could be a potential problem only for three systems, namely the
Perseus, Coma, and Virgo clusters.

Owing to its low Galactic latitude, the Perseus cluster is not a
member of the BCS. For Coma, VTP obtains a final flux of $3.19 \times
10^{-10}$ erg cm$^{-2}$ s$^{-1}$, in perfect agreement with the value
of $3.2 \times 10^{-10}$ erg cm$^{-2}$ s$^{-1}$ found by Briel, Henry
\& B\"ohringer (1992) who analyze the RASS data for Coma out to almost
twice the radius of our PET fields, namely 1.75 deg. The X-ray
emission from the Virgo cluster, however, is extended over more than
10 degrees on the sky; thus there can be no hope to recover its X-ray
flux completely from the data in our $2\times 2$ deg$^2$ field centred
on M87. VTP finds a total flux of $6.7 \times 10^{-10}$ erg cm$^{-2}$
s$^{-1}$, some 37 per cent of the total flux measured by B\"ohringer
et al.\ (1994) in an analysis of the RASS data in a $12\times 12$
deg$^2$ field.  We correct for the flux outside our PET field by
scaling the VTP count rate, flux and luminosity of Virgo accordingly,
i.e.\ by a factor of 2.74.

\section{Removal of non-cluster sources}
\label{clean_up}

The tentative sample of 1183 RASS-VTP sources described in
Section~\ref{ten_sample} was compiled with the goal of maximal
completeness. We included X-ray sources out to considerable distances
from the optical centroids of Abell and Zwicky clusters as potential
cluster detections, and, for the X-ray selected systems, accepted VTP
sources with extent values as low as 15 arcsec. The price to be paid
for the high degree of completeness thus achieved is severe
contamination by non-cluster sources. A large part of the
contamination is caused by random coincidences between the optical
Abell/Zwicky cluster positions and non-cluster X-ray sources, but also
by true coincidences with point sources like QSOs or AGN which are
members of otherwise X-ray faint clusters of galaxies.  Besides that,
our tentative sample will also contain a substantial number of
apparently extended VTP sources that are in fact blends of fainter
sources that can not be identified as such from the X-ray images
alone.

In order to identify chance coincidences as well as cluster-related
point sources, we cross-correlate our source list with the objects in
the NASA/IPAC Extragalactic Database (NED). We also obtain $12' \times
12'$ optical images from the digitized Palomar Observatory Sky Surveys
(POSS, first and second generation where available) using the {\sc
SkyView} and {\sc STScI} WWW interfaces\footnote{\tt
http://skview.gsfc.nasa.gov/skyview.html\\
http://archive.stsci.edu/dss} and look for possible optical
identifications. Wherever possible, we obtain E magnitudes and O--E
colours from the APM object catalogue (Automated Plate Measuring
machine, Irwin, Maddox \& McMahon 1994) for galaxies or stellar
objects thought to be the true source of the observed X-ray emission.

Also, we searched the ROSAT public archive of pointed PSPC and High
Resolution Imager (HRI) observations for further X-ray data on our
cluster candidates.  The deeper X-ray images thus obtained plus a few
more from non-public pointed observations awarded to the authors were
then examined in order to ensure that the emission detected by VTP in
the RASS data is genuinely extended indicating that it originates
indeed from a diffuse ICM.

We emphasize that, although we will loosely refer to the process
described in the following as the identification procedure, it is not
our goal to completely identify all X-ray sources in our tentative
sample. The only distinction we attempt to make for all sources is the
one between X-ray detections of diffuse emission from clusters of
galaxies and X-ray sources for which the emission is of different
origin. Although this step often entails identifications if the
optical non-cluster counterpart is either obvious or well-known, there
are cases where we discard sources as not cluster related without
having a secure alternative identification. The majority of these are
apparently extended, faint X-ray sources whith positions on the sky
that do not coincide with any discernible galaxy concentration but
that have at least one plausible stellar optical counterpart that
could account for part of the X-ray emission. Such sources are
consequently classified as likely blends of faint point sources.

Prior to assessing the available data on all our cluster candidates,
we need to establish a value for the radius of the error circle around
the nominal VTP X-ray source position within which to search for
optical counterparts. A good estimate of the intrinsic accuracy of the
VTP source positions can be obtained by cross-correlating them with
the sources of the RASS 1RXS catalogue (also know as Bright Source
Catalogue, Voges et al.\ 1996) which is a compilation of RASS sources
found by the standard SASS point source detection algorithms in the
merged RASS photon data (see also Section~\ref{rassbsc}). For bright
point sources, the 1RXS positions are expected to be accurate to about
10 arcsec of which 6 arcsec are assumed to be due to systematics. From
the cumulative distribution of the separations between VTP sources and
their 1RXS counterparts we derive a 90 per cent error radius of 1
arcmin around the positions of VTP point sources. For VTP point
sources with actual counterparts in the 1RXS sample we use the more
accurate source positions from the latter sample and adopt an error
circle of 30 arcsec radius in the identification process. For extended
VTP sources, we find significantly larger separations between the VTP
and the 1RXS positions. However, in general it is not clear which of
the two data sets can be considered to be the more accurate: SASS uses
a source profile that is inappropriate for extended emission and
likely to bias the source position toward the most pointlike feature
within the emission region. VTP on the other hand tends to blend
sources if both the detection and count rate thresholds are low. For
our present purpose of discarding non-cluster sources we take a
conservative approach and adopt a value of 2 arcmin for the search
radius around the positions of extended VTP sources. VTP extended
sources that have pointlike counterparts in the 1RXS sample within 2
arcmin are treated as potential blends (not all components of which
have necessarily been detected by SASS) if likely optical counterparts
are found within 30 arcsec of these pointlike 1RXS sources.

In the following we derive criteria designed to discriminate between
clusters and non-cluster sources using the X-ray properties of the VTP
detections and the optical properties of the most likely APM
counterparts within the X-ray error circle. We stress that although a
preliminary decision is made based on these criteria, no source is
discarded from our list of cluster candidates without careful visual
inspection of all the data at hand.

A brief summary of this procedure covering all steps detailed in
Sections~\ref{clean_up} and \ref{cr2flx} is given in
Table~\ref{clean_table}.

\begin{table}
\begin{tabular}{r@{\hspace{5mm}}p{0.38\textwidth}}
 number of   \hspace*{-3.5mm} &        \\
 VTP sources \hspace*{-5mm} & \hspace*{25mm} action taken\\ \hline \\
 1183          & {\sc Start:} initial list of candidate clusters\\ \hline 
 $-\;\;111$    & excluded as non-clusters based on pointed data \\
 $-\;\;\;\;71$ & excluded as catalogued AGN or QSOs without pointed data\\
 $-\;\;\;\;83$ & excluded as catalogued bright stars without pointed data\\
 $-\;\;\;\;62$ & excluded as likely stars (stellar object with E$\le 12$ 
                 within error circle) \\
 $-\;\;306$    & excluded as likely AGN or QSOs (object with E$\ge$12, O--E<2
                 within error circle)\\
 $-\;\;\;\;88$ & excluded as too soft (X-ray hardness ratio below 95 per cent
                 confidence limit for clusters)\\
 $+\;\;\;\;30$ & brought back as likely clusters after visual inspection of
                 all data at hand (overrides previous, automatic assessment) \\
 $-\;\;\;\;24$ & excluded as unlikely clusters after visual inspection of
                 all data at hand (overrides previous, automatic assessment) \\
 $-\;\;\;\;\;\;1$ & excluded (Z\,569 -- cluster but just south of $\delta=0^{\circ}$) \\
 $-\;\;\;\;\;\;1$ & excluded (unrelated source erroneously associated with Z\,5948) \\
 $-\;\;\;\;\;\;1$ & excluded as a blend of fainter sources after visual
                    inspection of all data at hand (overrides previous, automatic
                    assessment) \\
 $+\;\;\;\;14$ & included as faint (count rates of less than 0.07 ct s$^{-1}$) 
                 fragments of clearly extended main cluster emission\\
 $-\;\;\;\;22$ & fragments merged with main body of emission from 15 different clusters\\
 $-\;\;\;\;\;\;2$ & excluded as below the 0.07 count s$^{-1}$ count rate
                    threshold even after merging of fragments\\
 $+\;\;\;\;\;\;1$ & A\,625b included to join A\,625a although count rate is below 
                    0.07 count s$^{-1}$ \\
 $-\;\;\;\;14$ & excluded as underluminous ($L_{\rm X} \mbox{ (0.1--2.4 keV)} 
                 < 5\times 10^{42}$ erg s$^{-1}$) \\ \hline
  442          & {\sc End:} final list of likely clusters \\
\end{tabular}
\caption[]{Overview of the `identification' procedure used to
           discriminate between clusters and non-cluster X-ray sources
           starting from our initial list of 1183 VTP sources selected
           as cluster candidates and ending at our final sample of 442
           likely clusters. See Sections~\ref{clean_up} and
           \ref{cr2flx} for details.}
\label{clean_table}
\end{table}

\subsection{Sources with pointed ROSAT data and the significance of
local galaxy overdensities}
\label{clean_pointings}

We begin the identification process by examining the 246 sources for
which pointed ROSAT data are available. Of these 246, we confirm 135
as clusters on the grounds that their X-ray emission is genuinely
extended in the pointings. More than 83(93) per cent of these 135
sources feature RASS-VTP extents in excess of 30(15) arcsec, and all
of them are coincident with obvious galaxy overdensities on the
optical plates. We attempt to make the latter statement more
quantitative with the help of the APM galaxy counts in the vicinity of
the X-ray source (we use the magnitudes and number counts from the E
plates in all of the following). We compute the surface densities of
APM galaxies and blends in circles of 2.5, 5, and 7.5 arcmin radius
around the VTP source positions down to 18th and 19th magnitude
respectively, and compare them to the mean surface density of galaxies
on the respective optical plate. The statistical significance of any
overdensity is then determined using Poisson statistics, and the
highest of the six significances (for two limiting magnitudes in each
of three apertures) measured for each source is adopted as the final
value. We find more than 81 per cent (110 out of 135) of the clusters
confirmed in pointed observations to feature galaxy overdensities that
are significant at the greater than $2\sigma$ level. The 25 clusters
for which the APM data yield a less significant galaxy overdensity (or
none at all) meet one or more of the following criteria: they are
either (a) very nearby ($z\la 0.05$) groups or poor clusters which
often extend over an area greater than our largest aperture, or (b)
richer systems at redshifts greater than 0.2 with a considerable
fraction of the cluster galaxies fainter than 19th magnitude where the
APM galaxy counts begin to become unreliable for the shallowest
plates, or (c) they fall on plates for which the mean galaxy surface
density was not available and had to be estimated.  Although the
measurement of galaxy overdensities based on the APM object catalogue
can thus support optical cluster identifications, it cannot replace
visual inspection of the POSS plates.

This becomes even more evident when the false-alarm rate is
considered: more than a third of the 111 VTP sources identified as
non-clusters based on the pointed data feature APM galaxy
overdensities that are significant at the greater than $2\sigma$
level. This had to be expected though: not only do star-galaxy
misclassifications and the presence of spurious APM sources in the
vicinity of, for instance, bright stars boost the apparent galaxy
overdensities around non-cluster sources; more importantly, most of
our PET fields are, by design, centred on known clusters of galaxies,
making chance superpositions between non-cluster X-ray sources and
regions of increased galaxy surface density more likely than in
randomly selected fields. A high APM value of the local galaxy
overdensity is thus neither a sufficient nor a necessary condition for
a cluster identification.

We therefore use the APM galaxy overdensity measurements only as one
of several statistical indicators to eliminate non-cluster sources:
VTP sources with a galaxy overdensity significant at the less than
$2\sigma$ level have a probability of only $\sim 20$ per cent of being
clusters of galaxies.

\subsection{APM colours and magnitudes of confirmed clusters 
            and catalogued AGN, QSOs, and stars}
\label{clean_agnstars}

In this next step we investigate the distribution of the O$-$E
($\approx B-R)$ colours and E ($\approx R$) magnitudes of the
brightest plausible optical counterpart in the error circles around
VTP sources with cluster and non-cluster identifications.

Using all the information at hand, we identify 111 VTP sources with
catalogued AGN and QSOs. Since 40 of these 111 have already been
flagged as non-cluster sources on the basis of their appearance in
pointed ROSAT observations (see Section~\ref{clean_pointings}), the
total number of sources identified as non-clusters up to this point is
182. For 83 per cent of the sources identified as AGN and QSOs the
tentative optical counterpart is bluer than O$-$E=2 and fainter than
E=12 while the same is true for only less than four per cent of the
confirmed clusters.
 
Finally, we search for catalogued bright stars as possible optical
counterparts for our VTP detections and accept 92, 85 (92 per cent) of
which are brighter than E=12. With nine of these 92 having previously
been flagged as non-cluster sources based on pointed data, the total
number of identified non-cluster sources rises to 265. We note that
the probability of a chance alignment within 30(60) arcsec between a
VTP source and a star of APM magnitude E=12 or brighter is less than
1(3.4) per cent, so that the risk of an erroneous identification of a
VTP source with a bright star is very small.

\subsection{The hardness ratio distribution of VTP sources}
\label{clean_hr}

Using again as a training set the sample of 135 clusters confirmed by
ROSAT pointed observations, we establish a 95 per cent lower limit to
the spectral hardness ratio (see EVB for details) as a function of the
equivalent Galactic column density of neutral hydrogen (Stark et
al. 1992). While more than half of the 265 sources discarded so far as
non-clusters fall below this hardness ratio threshold\footnote{given
by $-0.39+0.83\log n_{\rm H}$ with $n_{\rm H}$ in units of $10^{20}$
cm$^{-2}$}, the same is, by design, true for only five per cent of the
confirmed clusters.

\subsection{Removal of unidentified non-cluster sources on the grounds
 of their X-ray/optical properties}

We now apply the criteria defined above to remove from our list of
cluster candidates all sources which are likely to be of non-cluster
origin. In addition to the 265 sources eliminated as point sources or
catalogued AGN, QSOs, and stars as described in
Sections~\ref{clean_pointings} and~\ref{clean_agnstars}, we

\begin{itemize}
 \item flag 62 sources as stars using the criterion defined in 
       Section~\ref{clean_agnstars} (stellar counterpart brighter than
       E=12 within the error circle),
 \item mark another 306 sources as likely AGN or QSOs using the criterion
       defined in Section~\ref{clean_agnstars} (E$\ge$12, O--E<2),
 \item and finally flag another 88 sources whose X-ray hardness ratio
       falls below the 95 per cent lower limit for clusters (see Section~\ref{clean_hr}).
\end{itemize}

Altogether we thus mark 456 X-ray sources as unlikely to be of cluster
origin on statistical grounds. However, since none of the applied
exclusion criteria is strictly {\em sufficient}, we expect some five
per cent, or 23, of these 456 sources to be clusters, despite the fact
that they lie close to a bright star, contain a very blue central
galaxy, or feature an unusually low X-ray hardness ratio. After
careful visual inspection of all tentatively discarded sources we
override our previous non-cluster assessment for 30 sources that
appear to be bona fide clusters from their combined X-ray/optical
appearance, bringing the number of sources in our sample of likely
clusters back up to 492 ($=1183-265-456+30$).

As our elimination criteria do not constitute a {\em necessary}\/ set
of conditions for a source to be of non-cluster origin, there will
also be a number of sources in this list of 492 that are in fact not
clusters but, e.g. X-ray bright stars with E>12 or AGN with colours
redder than O-E=2. Statistically, we expect some 11 per cent, or 39,
of the 357 sources presently marked as likely clusters without
confirmation by pointed data to be of non-cluster origin. Again, we
scrutinize all available information and discard a further 24 sources
as non-clusters on the grounds that the optical/X-ray appearance of
the source rules out a cluster identification. Six of these 24 sources
are in fact spurious in the sense that they do not correspond to any
well defined photon overdensity; rather, they are VTP detections of
large-scale background variations that extend over a significant
fraction of the $2\times 2$ deg$^2$ PET field. Other sources either
have obvious non-cluster counterparts on the optical plates which,
however, fail to meet our exclusion criteria (e.g., very bright
isolated spiral galaxies), or feature conflicting X-ray and optical
properties in the sense that they are clearly pointlike at the
resolution of the RASS but have been associated with very nearby
clusters whose galaxy distributions extend over tens of
arcminutes. Three sources are eliminated based on deep CCD images
taken with the University of Hawaii's 2.2m telescope which fail to
show a galaxy overdensity down to an R magnitude of 23. We also
discard Z\,569 at this stage since it falls just outside our study
area.

In the course of the visual inspection of both the X-ray and the
optical images of all our candidate sources we came repeatedly across
very extended VTP detections that do not have any obvious counterparts
at all in the optical. The emission from these very diffuse sources
is extended over typically several tens of arcminutes and spectrally
soft which is why we excluded the corresponding sources from our
list. Note that although these RASS-VTP detections are not always
obvious even to the eye, they are not spurious.  RXJ0905.4+1731 is a
typical example of this type of source and is also detected as an
extremely diffuse source at the edge of the field of the PSPC pointing
on the quasar 0903+169. This is the same type of source as the one
detected first in the RASS by Ebeling et al.\ (1994) who tentatively
classified them as fragments of Galactic supernova remnants.

The visual inspection of all our candidates also causes one further
detection of an Abell cluster to be included in our sample. A\,1758b
is more than 0.470 Abell radii away from Abell's optical cluster
position and was thus not included in the sample selected from the
cross-correlation between the Abell catalogue and all X-ray sources
detected by VTP. However, it is picked up later as a possibly extended
VTP source (with an extent parameter of 17 arcsec) and joins
A\,1758a. Another source, initially included because it falls within
9.1 arcmin of Z\,5948, is eliminated at this stage on the grounds that
it is pointlike in the VTP analysis and clearly unrelated to Z\,5948
($=$A\,1738) which is detected as a different VTP source at the
nominal Abell position.

\begin{figure}
  \epsfxsize=0.5\textwidth
  \hspace*{0cm} \centerline{\epsffile{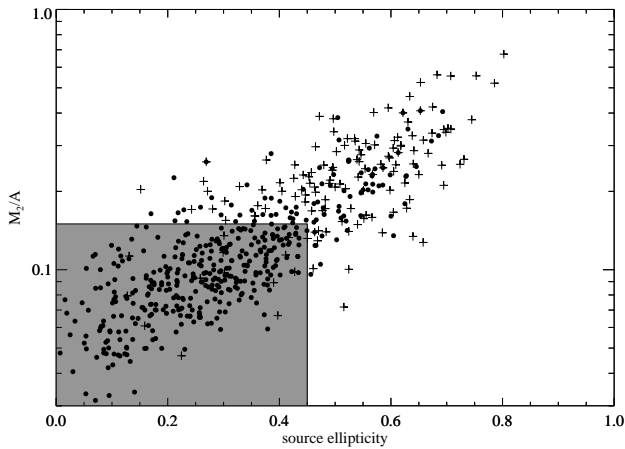}}
  \caption[]{Second moments ($M_2$) over source area ($A$) versus source
           ellipticity $e$ for the VTP detections classified as blends
           (crosses) and clusters (filled circles). Source ellipticity
           is defined as $e=1-b/a$ where $b$ and $a$ are the source's
           minor and major axes, respectively. Note that more than 90
           per cent of all blends lie outside the shaded region. }
           \label{vtp_blend_cand}
\end{figure}

As mentioned earlier, blended sources become increasingly common as
the flux of the sources in question approaches the detection limit. In
an attempt to find blended sources that might have been missed by the
above screening procedure we investigate the efficiency of two more
source parameters as an additional criterion to discriminate against
blends.  Fig.~\ref{vtp_blend_cand} shows the ratio of the second
moment of the sources' spatial photon distribution, $M_2$, over the
area covered by the source photons, $A$, as a function of source
ellipticity $e$.  Whereas the sources classified as clusters extend
over a considerable range of values in both parameters, the sources
eliminated so far as likely blends are preferentially found outside
the shaded region defined by $e \le 0.45$ and $M_2/A \le 0.15$ in
Fig.~\ref{vtp_blend_cand}.  We therefore investigate once more the 94
sources classified as clusters that fall into the realm of blends
above the quoted thresholds\footnote{Note that these empirical
threshold values have no theoretical foundation but are solely based
on the distribution of blends and clusters shown in
Fig.~\protect\ref{vtp_blend_cand}.}.

Reassuringly, the re-investigation of these sources does not lead to a
substantial change in our initial assessment: only one source
previously classified as a cluster is now discarded as a likely blend
with a dominant non-cluster component. Essentially, the clusters in
the high ellipticity/high $M_2/A$ regime can be divided into three
categories: clusters with good X-ray photon statistics that are indeed
distinctly non-spherical (like A76, A1314 or A2111), sources with poor
photons statistics whose ellipticity and $M_2/A$ values are ill
determined or whose VTP detections contain shot noise from the
vicinity of the source causing the emission to appear more elliptical
than it is, and, finally, clusters whose emission is blended with
X-rays from other sources that may or may not be part of the cluster
but appear more peaked than the diffuse cluster emission. Most of the
clusters in the third category have in fact already been classified as
possibly contaminated in our initial inspection, the remainder receive
their `contaminated?' flag now.

Altogether we have thus removed 718 X-ray sources from our initial
candidate list of 1183 VTP detections which leads to a cleaned sample
of 465 X-ray sources. Although below the count rate limit of 0.07
count s$^{-1}$, a further 14 VTP detections are added to the source
list as fragments of extended cluster emission that has been detected
in several pieces by VTP. Like 23 brighter VTP sources already
included in our list under the same classification, these detections
do not appear to define any obvious substructure in the X-ray emission
from the 15 different clusters affected. Following EVB, we
therefore merge the multiple detections into one for each of these 15
thereby reducing our VTP source list by 22 entries. Two of the
multiply detected clusters fall short of our count rate limit even
after the individual components have been merged and are thus rejected
which leaves us with 455 VTP detections of 455 clusters. Seven of
these are subclusters (A\,625a, A\,1758a/b, A\,2197a/b, and
A\,2572a/b) detected as separate entities by VTP. The one subcluster
missing to complete the above list (A\,625b) was initially excluded
because its VTP count rate lies below 0.07 count s$^{-1}$. We include
it now which brings the number of clusters in our sample to 456, four
of them double.

\section{Converting PSPC count rates into fluxes}
\label{cr2flx}

The conversion of our final corrected VTP count rates into energy
fluxes follows again EVB.

We convert the corrected count rates to proper, unabsorbed energy
fluxes using the XSPEC spectral analysis package and assuming a
Raymond-Smith type spectrum with a global metal abundance of 30 per
cent of the solar value\footnote{We use the solar abundance table of
Anders \& Grevesse (1989)}. Values for the Galactic column density of
neutral Hydrogen, $n_{\rm H}$, are taken from the compilation of Stark
et al.\ (1992). Where available, measured X-ray temperatures taken
from the compilation of David et al.\ (1993) are used in the
conversion; for the remainder the ICM gas temperature is estimated
from the clusters' bolometric X-ray luminosity. We adopt
\[ 
{\rm k}T = 2.76\,{\rm keV}\, L_{\rm X,bol,44}^{0.33} 
\] 
for the k$T-L_{\rm X}$ relation (White et al.\ 1997) where $L_{\rm
X,bol,44}$ is the bolometric X-ray luminosity in units of $10^{44}$
erg s$^{-1}$. Starting from a default value of 5 keV, k$T$ is then
determined from the above relation in an iteration loop. K corrections
are applied using measured cluster redshifts wherever available (357
clusters) and estimates for the remainder of the sample (99 clusters).

In addition to computing fluxes and cluster-rest-frame luminosities in
the ROSAT broad band ($0.1-2.4$ keV) and the pseudo-bolometric
$0.01-100$ keV band, we also determine the respective values for all
of our clusters in the energy ranges of previous missions, namely
$0.5-2.0$ keV (ROSAT hard band), $0.3-3.5$ keV (EMSS), and $2-10$ keV
(HEAO-1) in order to facilitate the comparison of our results with
that of previous and future studies.

Using the derived X-ray luminosities in the ROSAT broad band, we apply
one more cut to our sample and exclude all `clusters' featuring X-ray
luminosities below $5 \times 10^{42}$ erg s$^{-1}$ ($0.1-2.4$ keV).
Most of these nearby, low luminosity systems have entered our sample
as purely X-ray selected sources classified as extended by either the
SASS or VTP. A total of 14 systems are thus eliminated from our list
(among them M31 and M82); one source, A\,2197b, is kept despite its
low luminosity in order to allow us to keep track of all components of
multiple systems. This last elimination step leaves us with our final
sample of 442 clusters\footnote{Eight of these systems are in fact the
components of four double clusters.}.

\section{The cluster $\log N-\log S$ distribution and the BCS flux limit}

Given the rather complicated compilation procedure (different source
detection algorithms with different sky coverage, a combination of
optical and X-ray criteria in the selection of our initial PET field
targets etc.), one might think that the completeness of the BCS would
be difficult to quantify.

Fortunately, this is not the case.

\subsection{Inclusion of serendipitous cluster detections}
\label{serdet}

\begin{figure}
  \epsfxsize=0.5\textwidth
  \hspace*{0cm} \centerline{\epsffile{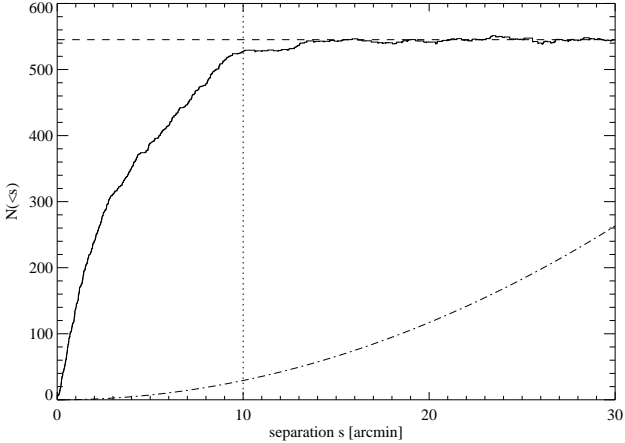}}
  \caption[]{The cumulative number of VTP sources with detected count
 	   rates higher than 0.07 count s$^{-1}$ from all 631 PET
 	   fields as a function of their angular distance from the
 	   respective field centres. A parabolic component (the
 	   dot-dashed line) describing the background source density
 	   of 0.53 deg$^{-2}$ has been subtracted. The fact that the
 	   data reach the constant background level (marked by the
 	   dashed line) only at separations of about 13 arcmin is due
 	   to peculiarities introduced by the `SASS crashes' discussed
 	   in Section~\protect\ref{sass_crash}.}
 	   \label{cumnum_serendipitous}
\end{figure}

Thanks to having VTP run on all of our 631 $2 \times 2$ deg$^2$ PET
fields, we have a complete set of {\em serendipitous}\/ VTP detections
that is almost totally independent of the initial selection criteria
applied to the SASS source list. The total sky area covered by all PET
fields is 2,190 deg$^2$, not much less than $631 \times (1.8 \,
\mbox{deg})^2 = 2,278$ deg$^2$ due to a small amount of overlap
between some of our fields. (We remind the reader that the outer 5 per
cent of each PET field are actually not processed by VTP but serve as
a tessellation margin.) Plotting the cumulative number of all VTP
detections with raw count rates in excess of 0.07 count s$^{-1}$ in
these 631 fields as a function of their radial separation from the
field centre, we find the number of sources per unit area (and above
the quoted count rate limit) to converge to 0.53 deg$^{-2}$ at a
radius of about 10 arcmin (see Fig.~\ref{cumnum_serendipitous}).  We
thus subtract another $631 \times \pi \times (10 \,\mbox{arcmin})^2 =
55$ deg$^2$ from the above 2,190 deg$^2$ to allow for the central
target regions of each field within which detections cannot be
considered to be serendipitous, and obtain a total solid angle of
2,134.6 deg$^2$ (or 15.7 per cent of the northern extragalactic sky)
that is available for serendipitous VTP detections. We note that no
additional corrections for obscuration by extended emission are
required: the total area covered by cluster emission detected by VTP
outside the mentioned inner circle of 10 arcmin radius in each field
is 1.7 deg$^2$, i.e., a negligible fraction of the total solid angle
available for serendipitous detections.

However, not all of the 100 cluster detections in these 2,134.6
deg$^2$ are necessarily serendipitous. ``Clusters cluster'' to the
extent that the odds of finding a second cluster within some 20 Mpc of
any given cluster are significantly increased over the probability
derived from their average volume density. As, by design, our
serendipitous detections are all within 1.3 deg of some target cluster
(i.e., within 20 Mpc for $z<0.2$) we need to account for the increase
in the projected surface density of clusters caused by the
non-vanishing amplitude of the cluster-cluster correlation function on
these scales.

\begin{figure}
  \epsfxsize=0.5\textwidth
  \hspace*{0cm} \centerline{\epsffile{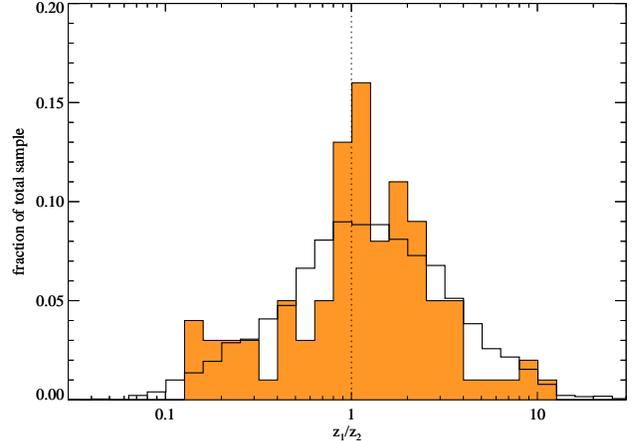}}
  \caption[]{The distribution of the redshift ratios of the 100
           serendipitous/target cluster pairs (shown shaded). The same
           distribution for a control sample of 5000 random pairs of
           target clusters separated by more than 20 deg on the sky is
           shown as the solid, bold line.}  \label{bcs_serend_hist}
\end{figure}

To this end we compare the relative separation in redshift space
between the serendipitous clusters and the respective closest target
clusters with the relative separation in $z$ between random pairs of
target clusters and serendipitously detected clusters that are
separated by more than 20 deg on the sky. The latter set of pairs
serves as a control sample which, owing to the large angular
separation within each pair, can be assumed to be unaffected by
clustering. As can be seen in Fig.~\ref{bcs_serend_hist}, the fraction
of serendipitous clusters that lie within some 20 percent of the
redshift of the nearest target cluster is somewhat higher than what
would be expected from the same distribution for physically unrelated
cluster pairs, indicating that clustering has indeed some
effect\footnote{The distribution shown in Fig.~\ref{bcs_serend_hist}
is slightly skewed toward $z_1/z_2$ values greater than unity. This is
due to a small, systematic difference in the redshift distributions of
the two samples: since we processed the PET data around {\em all}\/
Abell clusters within a redshift of 0.05, the odds of serendipitously
detecting a background cluster are, on average, slightly higher than
those of finding a cluster in the foreground.}.  Note that the fact
that more than half of the serendipitously detected clusters have only
estimated redshifts does not greatly affect this result: Although
random errors in the estimated redshifts (which are of the order of 20
per cent, see EVB) introduce a blurring of the distribution shown in
Fig.~\ref{bcs_serend_hist}, the effect is limited to a scale
comparable to the bin width in Fig.~\ref{bcs_serend_hist}. We can, in
a statistical sense, quantify how many of our serendipitously detected
clusters can be attributed to large-scale clustering by gradually
removing from our sample quasi-serendipitous clusters that lie close
to the peak of the distribution shown in Fig.~\ref{bcs_serend_hist}
until the redshift ratio distribution of the remaining cluster pairs
becomes consistent with that of our control sample. In any case we are
dealing with a small correction as, according to the
Kolmogorov-Smirnov statistic, the distribution of the original sample
is inconsistent with that of the control sample at only the 56 per
cent level. If only five serendipitous clusters that lie within 20 per
cent of the redshift of the nearest target cluster are removed, the
probability that the remaining distribution is not drawn from the same
parent distribution as that of our control sample falls to below 50
per cent, and if 10 of the serendipitous detections are attributed to
clustering, the respective figure drops to less than 23 per cent.

In an attempt to find at least some of the few physical pairs, we
compute the projected metric separations of those serendipitous/target
cluster pairs whose redshifts agree to within 20 per cent. Five pairs
feature separations of less than 2 Mpc if the mean of the respective
cluster redshifts is assumed in the conversion from angular to metric
separations. We assume that the affected serendipitously detected
clusters are physically related to the respective target clusters, and
thus move A\,2122, A\,2151b, A\,2197b, A\,2678, and IV\,Zw\,038 from
our list of serendipitous detections to that of our target
clusters. Note that all of these five clusters have measured
redshifts.

\begin{figure}
  \epsfxsize=0.5\textwidth \hspace*{0cm} \centerline{\epsffile{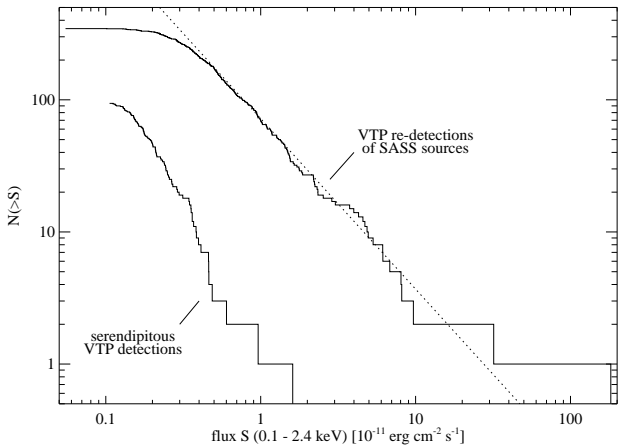}} 
  \caption[]{The cumulative flux distribution of the 442 clusters in our
             sample of VTP detections with raw count rates greater
             than 0.07 count s$^{-1}$. VTP re-detections of the
             initial target clusters selected from the SASS source
             list and additional, serendipitous VTP detections are
             plotted separately. The sample of SASS-based clusters
             also contains five physically associated systems. Note
             how, at a flux of about $5\times 10^{-12}$ erg cm$^{-2}$
             s$^{-1}$, the number count distribution of SASS-based
             clusters begins to flatten compared to a power law (shown
             as a dotted line).}  \label{raw_lognlogs}
\end{figure}

The set of the remaining 95 serendipitous VTP detections (of which
no more than a handful may still be due to clustering) is of crucial
importance for the assessment of the completeness of the BCS as a
statistical sample. As mentioned before, VTP was run on only 15.7 per
cent of our actual study area. Hence, the BCS cluster list is
incomplete, to some extent, at almost any X-ray flux level, and
particularly so for low flux and low surface brightness emission from
nearby poor clusters and groups of galaxies that the SASS detection
algorithms are particularly insensitive to. This is illustrated in
Fig.~\ref{raw_lognlogs} which shows the cumulative number of clusters
in our (so far) VTP count rate limited sample as a function of energy
flux with the total sample split into two subsets.  On the one hand,
there are the 342 clusters detected by VTP within the inner 10 arcmin
of the PET fields (which are, with few exceptions, re-detections of
sources from our initial SASS source list which covered the whole
northern extragalactic sky) plus 5 clusters detected at larger angular
offsets from the respective target clusters but physically associated
with them. On the other hand, we have the 95 clusters detected
serendipitously by VTP in our 631 PET fields (i.e. outside the inner
10 arcmin -- again with a few exceptions).  Note how the sample based on
SASS detections alone starts to become seriously incomplete at fluxes
as high as $\sim 5\times 10^{-12}$ erg cm$^{-2}$ s$^{-1}$.

The level of incompleteness at any flux level can be quantified very
accurately with the help of the serendipitous detections.  The total
solid angle of our study area defined by $\delta \geq 0^{\circ}$, $|b|
\geq 20^{\circ}$ is 13,578 deg$^2$, or 4.14 sr. By weighting the
serendipitous clusters by a factor given by the ratio of respective
sky areas (i.e., $13,578/2,134.6 = 6.36$), we can thus correct for the
incompleteness introduced by the fact that VTP was not run on the
whole of our study area. This would be the correct weight only if our
cleaned list of serendipitous detections was really entirely
unaffected by clustering. If we assume that five more of our 95
non-target clusters are in fact not really serendipitous (without
knowing which ones they are), the weight assigned to all our
non-target clusters drops from
\[
   w_0 = \frac{13,578}{2,134.6} = 6.36 \quad 
             \mbox{to} \quad w = w_0 - \frac{5}{95}\,(w_0-1) = 6.08.
\]

For many statistical purposes (such as studies of the cluster $\log
N-\log S$ distribution or the cluster X-ray luminosity function) the
intrinsic incompleteness of the BCS can thus be entirely overcome by
applying the above modified sky coverage correction. This is
demonstrated in the following section.

\subsection{The $\log N-\log S$ distribution of clusters of galaxies}

As was shown in the previous section, the BCS cluster list starts to
become seriously incomplete at fluxes around $5\times 10^{-12}$ erg
cm$^{-2}$ s$^{-1}$. However, when the serendipitous VTP detections are
normalized properly, it becomes apparent that this incompleteness is
not due to any intrinsic limitation of the RASS like, for instance,
the relatively short exposure times, but entirely to the inadequacy of
the SASS algorithms for the detection and characterization of extended
emission.

\subsubsection{The BCS $\log N-\log S$ distribution}
\label{logNlogS}

\begin{figure}
  \epsfxsize=0.5\textwidth
  \hspace*{0cm} \centerline{\epsffile{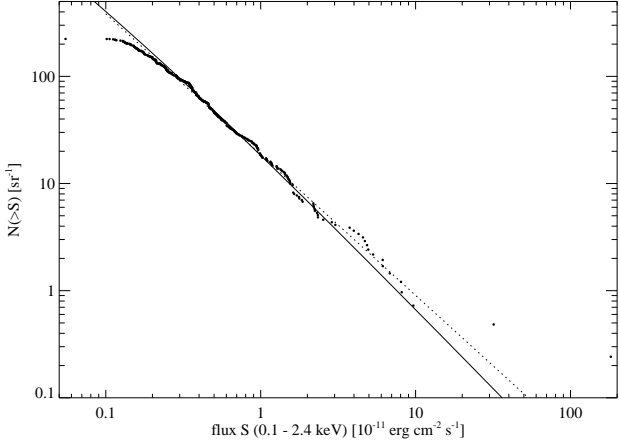}}
  \caption[]{The cumulative $\log N-\log S$ distribution of the 442
           clusters in our sample of VTP detections with raw count
           rates greater than 0.07 count s$^{-1}$. Serendipitous VTP
           detections are included with a relative weight of 6.08
           which takes into account the smaller solid angle available
           for serendipitous detections. The solid curve shows the run
           of $N(>S)$ for a cluster population described by the BCS
           X-ray luminosity function (Ebeling et al.\ 1997) in a flat
           Einstein-de Sitter Universe. The dotted line represents the
           best overall power law description of the data and has a
           slope of $-1.31$.}  \label{lognlogs_ros}
\end{figure}

Fig.~\ref{lognlogs_ros} shows the BCS $\log N-\log S$ distribution
when serendipitous VTP detections are included with a relative weight
of 6.08 which allows for the smaller solid angle that is actually
available for serendipitous detections (see Section~\ref{serdet} for
details). We determine the best power law description of the BCS $\log
N-\log S$ in a maximum likelihood fit to the bright end of the
differential, unbinned distribution. As the uncertainty of the
best-fitting slope is dominated by systematics (namely the precise
choice of the flux range over which the fit is performed) rather than
by statistics, we investigate these systematics by performing a series
of fits to the data in the range $S>S_{\rm min}$, where $S_{\rm min}$
is varied from $2\times 10^{-12}$ to $2\times 10^{-11}$ erg cm$^{-2}$
s$^{-1}$. Note that, in these fits, the amplitude $\kappa$ of the
fitted power law $N(>S) = \kappa\, S^{-\alpha}$ is completely
determined by the best-fitting slope $\alpha$ and the requirement that
the integral number of clusters be conserved.  Thus, $\alpha$ is the
only independent fit parameter. For $S_{\rm min}\approx 9\times
10^{-12}$ erg cm$^{-2}$ s$^{-1}$ we find the BCS $\log N-\log S$
distribution to be well described by a Euclidean power law, i.e.\
$\alpha=1.5$, indicative of a non-evolving source population in a
static Universe. For lower values of $S_{\rm min}$ the best-fitting
$\alpha$ value decreases slightly to reach $\alpha\approx 1.3$ at
$S_{\rm min} = 2.5\times 10^{-12}$ erg cm$^{-2}$ s$^{-1}$. This
intrinsic flattening of the cluster $\log N-\log S$ distribution is,
however, not necessarily a sign of cluster evolution but mainly an
effect of the expansion of the Universe. At redshifts below about 0.1,
spatial volumes in the observer's reference frame can be considered
identical to the intrinsic, comoving ones. However, at the redshifts
probed by the BCS ($z\la 0.4$), the difference becomes significant.
If cosmological effects are taken into account, we find not just the
bright end, but in fact the whole of the BCS $\log N-\log S$
distribution to be fully consistent with a non-evolving cluster space
density. Fig.~\ref{lognlogs_ros} illustrates this by showing the run
of $N(>S)$ for a cluster population following the local cluster X-ray
luminosity function (XLF) as described by Ebeling et al.\ (1997) and
assuming an Einstein-de Sitter Universe where $q_0=0.5$ and $H_0=50$
km s$^{-1}$ Mpc$^{-1}$ (solid line). Although the good agreement is
not entirely surprising in view of the fact that both the $\log N-\log
S$ distribution and the cluster XLF were determined from the same
sample (the BCS), the comparison still represents a useful consistency
check.  Note also that the excellent agreement holds down to fluxes of
about $2.5\times 10^{-12}$ erg cm$^{-2}$ s$^{-1}$, i.e., almost a
factor of two below the flux limit of the BCS subsample used by
Ebeling et al.\ (1997) in the construction of the cluster XLF.

In order to facilitate comparison with previous work and also to
provide a simple analytic parameterization of the BCS $\log N-\log S$
distribution over a large range of fluxes, we determine a global power
law description from the values obtained for $\kappa$ and $\alpha$
when $S_{\rm min}$ is varied from $4\times 10^{-12}$ to $8\times
10^{-12}$ erg cm$^{-2}$ s$^{-1}$. The median values $\kappa, \alpha$
of the best fitting power laws found in this range are
\begin{equation}
	\kappa = 18.63^{-0.08}_{+0.09}\;
        \mbox{sr$^{-1}$ (10$^{-11}$ erg cm$^{-2}$ s$^{-1}$)$^\alpha$},
              \quad 
        \alpha = 1.31^{+0.06}_{-0.03}
 \label{lognlogspar}
\end{equation}
where the quoted errors do not represent $1\sigma$ uncertainties but
rather the $10^{\rm th}$ and $90^{\rm th}$ percentiles of the
distribution of best-fit parameters.

We emphasize once more that the fact that the Euclidean $\alpha$ value
of 1.5 is excluded at more than 90 per cent confidence by the quoted
uncertainties does {\em not}\/ imply an evolving cluster population.

\begin{figure}
  \epsfxsize=0.5\textwidth
  \hspace*{0cm} \centerline{\epsffile{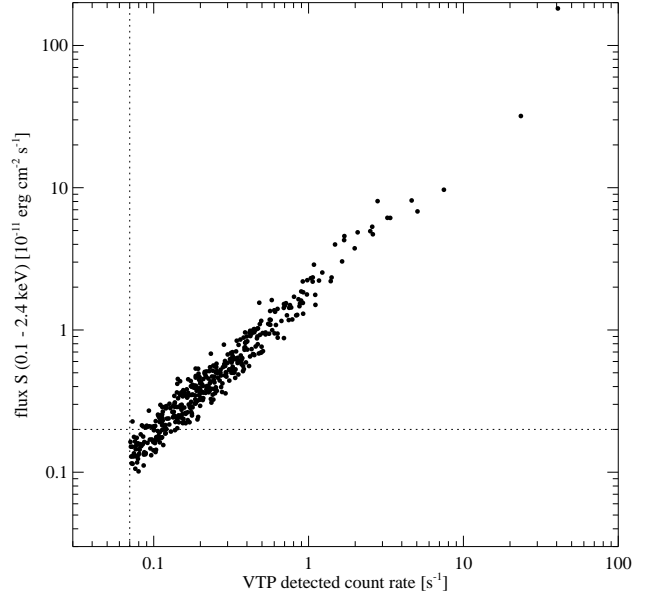}}
  \caption[]{The final cluster fluxes as a function of the detected VTP
           count rate for our sample of 442. Note how our count rate
           limit of 0.07 count s$^{-1}$ ensures completeness to a flux
           limit of about $2 \times 10^{-12}$ erg cm$^{-2}$ s$^{-1}$.}
           \label{rawcr_flux}
\end{figure}

It is also noteworthy that, at a flux of $2\times 10^{-12}$ erg
cm$^{-2}$ s$^{-1}$, the data still fall within four per cent of the
power law value of equation~\ref{lognlogspar} and within nine per cent
of the prediction of our Einstein-de Sitter model, indicating that, in
principle, the RASS data in the northern hemisphere would allow the
compilation of complete, flux-limited cluster samples down to, and
possibly even below, this value. At even lower fluxes the BCS $\log
N-\log S$ begins to flatten beyond the theoretical expectation. This
is, however, probably mainly due to our count rate limit at 0.07 count
s$^{-1}$ as can be seen from Fig.~\ref{rawcr_flux} which shows the
final cluster fluxes as a function of the initially detected (i.e.\
uncorrected) VTP count rate. At higher count rates (fluxes) VTP
detects more than 50 per cent of the cluster emission directly
regardless of the extent of the emission (cf.~Fig.~\ref{fc_rc}), i.e.\
the detection efficiency is indeed essentially 100 per cent.

Thanks to the, by X-ray standards, huge size of our sample (442
clusters) and the homogeneous selection criteria used in its
compilation, the power law with $\kappa$ and $\alpha$ as quoted in
equation~\ref{lognlogspar} represents the best parameterization of the
cluster $\log N-\log S$ distribution at fluxes above $2\times
10^{-12}$ erg cm$^{-2}$ s$^{-1}$ to date.

\subsubsection{Comparison with previous determinations of the cluster 
               $\log N-\log S$ distribution}

To allow a comparison of our results with previous and future work in
other energy bands, we also compute the BCS $\log N-\log S$
distribution in the following energy ranges: $0.5-2.0$ keV (ROSAT hard
band), $0.3-3.5$ keV (EMSS), $2-10$ keV (HEAO 1), and $0.01-100$ keV
(pseudo-bolometric).

\begin{figure}
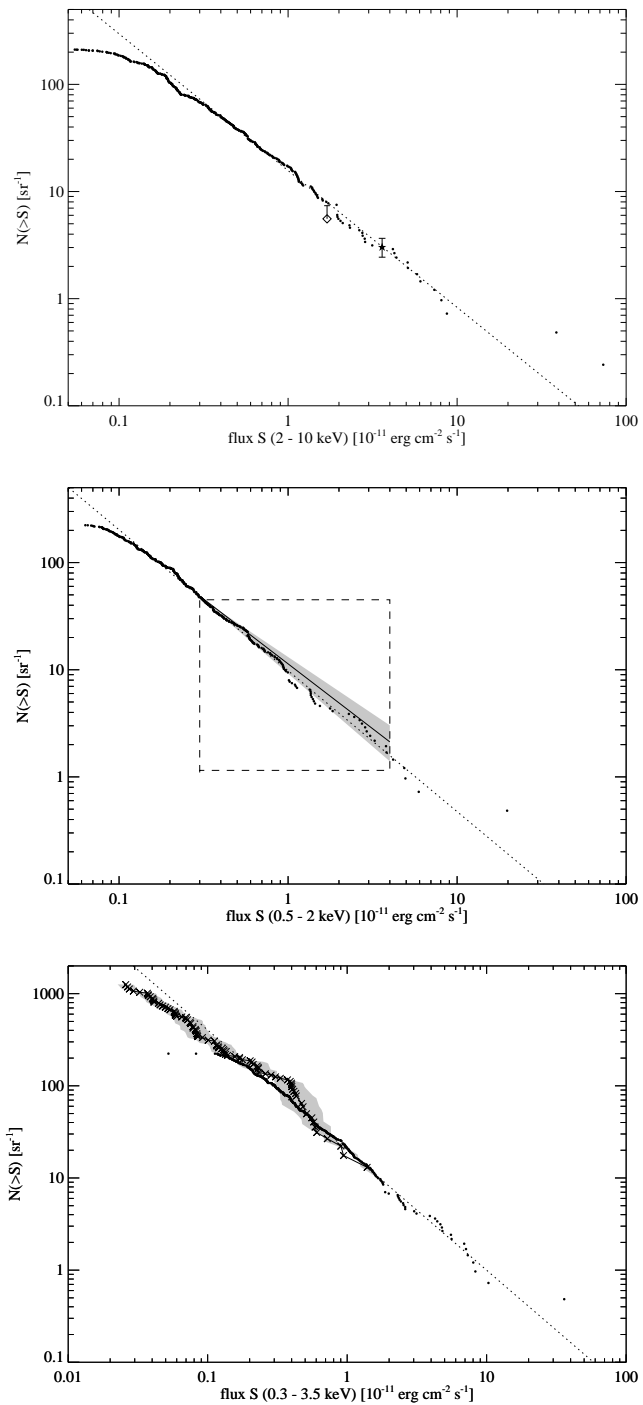

  \epsfxsize=0.5\textwidth
  \hspace*{0cm} \centerline{\epsffile{bcs_lognlogs_2-10.epsf}}
  \epsfxsize=0.5\textwidth
  \hspace*{0cm} \centerline{\epsffile{bcs_lognlogs_0.5-2.epsf}}
  \epsfxsize=0.5\textwidth
  \hspace*{0cm} \centerline{\epsffile{bcs_lognlogs_0.3-3.5.epsf}}
  \caption[]{The cumulative $\log N-\log S$ distribution of the 442
           clusters in our sample in the $2-10$ keV (top), the
           $0.5-2.0$ keV (centre), and the $0.3-3.5$ keV band
           (bottom).  The single data point shown as an asterisk in
           the top panel denotes the space density of HEAO-1 clusters
           as determined by Piccinotti et al.~(1982); the open diamond
           represents the result of Edge et al.\ (1990) with its 90
           per cent upper limit. On the central panel, the solid line
           inside the shaded triangle represents the best power law
           fit with its $1\sigma$ error range for the $\log N-\log S$
           distribution of De Grandi (1996) for an early cluster
           subsample from the ESO Key Programme (now known as the
           REFLEX Cluster Survey). The actual flux range covered by
           the REFLEX subsample is indicated by the dashed box.
           Bottom panel: The crosses connected by a solid line
           represent the $\log N-\log S$ distribution of the EMSS
           cluster sample of Henry et al.\ (1992). The shaded region
           illustrated the effect of varying the assumed cluster core
           radius from 200 to 300 kpc. At its bright end, the EMSS
           $\log N-\log S$ distribution was forced to agree with the
           BCS power law.}  \label{lognlogs_more}
\end{figure}

Fig.~\ref{lognlogs_more} shows the respective distributions in three
of these bands and illustrates the excellent agreement between our
results and earlier measurements. In the $2-10$ keV band
(Fig.~\ref{lognlogs_more}, top panel), at very high fluxes ($S>3.6
\times 10^{-11}$ erg cm$^{-2}$ s$^{-1}$), our data confirm the early
measurement of Piccinotti et al.\ (1982) which was based on 25
clusters detected by the HEAO-1 A-2 experiment. At somewhat lower
fluxes, the BCS $\log N-\log S$ is also (just) in agreement with the
result of Edge et al.\ (1990) obtained for a sample of 46 clusters
compiled from HEAO-1, ARIEL-V, EINSTEIN, and EXOSAT data.

In the $0.5-2.0$ kev range (Fig.~\ref{lognlogs_more}, middle panel),
at fluxes down to $3 \times 10^{-12}$ erg cm$^{-2}$ s$^{-1}$, the BCS
$\log N-\log S$ distribution is, within the errors, in agreement with
early results from the ESO Key Programme\footnote{recently renamed
REFLEX Cluster Survey} (Guzzo et al.\ 1995) for a sample of 111
clusters in the southern hemisphere (De Grandi 1996). That the
best-fitting power law found by De Grandi is somewhat shallower than
that of the BCS in the same flux range ($\alpha = 1.21$ as opposed to
1.28 for the BCS) is mainly due to the fact that De Grandi's fit does
not explore a range of minimal fluxes which causes her best power law
representation to be dominated by the apparent flattening of the $\log
N-\log S$ at the low flux end of her sample.  However, her $1\sigma$
lower limit ($\alpha = 1.36$, $\kappa = 9.21$) is in very good
agreement with both the BCS and the Piccinotti et al.\ result.

The bottom panel of Fig.~\ref{lognlogs_more}, finally, shows the BCS
$\log N-\log S$ distribution in the $0.3-3.5$ band of the EINSTEIN
observatory. To establish the EMSS cluster $\log N-\log S$
distribution, we consider the updated EMSS cluster sample of Gioia \&
Luppino (1994) with the restrictions described by Henry et al.\
(1992), i.e., $\delta \geq -40^{\circ}$ and $f_{\rm det} \geq 1.33
\times 10^{-13}$ erg cm$^{-2}$ s$^{-1}$ ($0.3-3.5$ keV). The latest
EMSS source identifications as listed by Maccacaro et al.\ (1994) are
taken into account as well as new redshifts from Luppino \& Gioia
(1995) and Romer (private communication). Also, we discard
MS$0011.7+0837$, MS$1050.7+4946$, MS$1154.1+4255$, MS$1209.0+3917$,
and MS$1610.4+6616$, all of which have revised non-cluster
identifications based on additional optical and X-ray observations
(Gioia, Stocke, Henry, private communication). The $\log N-\log S$
distribution for the resulting sample of 87 EMSS clusters is then
computed from the total cluster fluxes and the sky coverage values
corresponding to the detected fluxes as published by Henry et al.\
(1992). Consisting exclusively of serendipitous detections, the EMSS
sample is potentially intrinsically incomplete at its high flux end,
as the corresponding X-ray bright clusters were preferentially
selected as targets of EINSTEIN observations thereby making them
unavailable for serendipitous detection. We compensate for this effect
by forcing the EMSS $\log N-\log S$ distribution to match the BCS
power law value at the flux of the brightest EMSS cluster.  We stress
that this is only a minor adjustment as no more than one additional
cluster is required at $S\ga 1.4 \times 10^{-11}$ erg cm$^{-2}$
s$^{-1}$ to make the bright end of the EMSS $\log N-\log S$
distribution match the BCS power law value.

In addition to the usual statistical error, the EMSS $\log N-\log S$
function is affected by systematic errors introduced by the fact that
Henry and coworkers used a fixed value of 250 kpc for the cluster core
radius in their conversion from detected to total fluxes.  The impact
of this effect can be assessed by recomputing the total EMSS cluster
fluxes for slightly different core radii (we assume 200 and 300 kpc,
respectively). The shaded region in the bottom plot of
Fig.~\ref{lognlogs_more} illustrates the resulting uncertainty in the
EMSS $\log N-\log S$ distribution. Within these errors, the EMSS data
are fully consistent with the BCS $\log N-\log S$ distribution where
the flux ranges of the two samples overlap and the BCS is not affected
by the applied count rate cut (i.e.\ $S\ga 2\times 10^{-12}$ erg
cm$^{-2}$ s$^{-1}$). At fluxes around $1\times 10^{-12}$ erg cm$^{-2}$
s$^{-1}$, however, the EMSS $\log N-\log S$ distribution begins to
deviate significantly from the extrapolated BCS power law, possibly
indicating that cluster evolution becomes increasingly important at
the higher redshifts probed by the EMSS ($z_{\rm max}=0.82$). The
evidence for cluster evolution in the BCS is discussed in the context
of the BCS X-ray luminosity function (Ebeling et al., 1997); a
reassessment of the significance of the evolution in the EMSS sample
is discussed by Nichol et al.\ (1997).

Since the only parameterization of the EMSS cluster $\log N-\log S$
distribution in the literature (Gioia et al., 1984) is based on an
erroneous sky coverage calculation which did not treat extended
emission properly, we quote the best-fit power law parameters for the
EMSS cluster sample for reference:
\begin{equation}
	\kappa = 23.9^{-4.3}_{+2.7}\;
        \mbox{sr$^{-1}$ (10$^{-11}$ erg cm$^{-2}$ s$^{-1}$)$^\alpha$},
              \quad 
        \alpha = 1.13^{+0.11}_{-0.05}.
 \label{emss_lognlogspar}
\end{equation}
The quoted errors correspond to the $10^{\rm th}$ and $90^{\rm th}$
percentiles of the best-fit values found when the fit range is varied
from $S_{\rm min}=4\times 10^{-13}$ to $8\times 10^{-13}$ erg
cm$^{-2}$ s$^{-1}$. A core radius of 250 kpc was assumed in the
conversion from detected to total fluxes. Varying this parameter in
the range from 200 to 300 kpc affects mainly the amplitude of the
resulting best-fitting power law but hardly changes the slope. The
best-fit parameters of equation~\ref{emss_lognlogspar} change only
marginally if we fit to the EMSS data without enforcing a match with
the BCS power law at the bright end. We obtain
\[
	\kappa = 20.7^{-4.2}_{+2.9}\;
        \mbox{sr$^{-1}$ (10$^{-11}$ erg cm$^{-2}$ s$^{-1}$)$^\alpha$},
              \quad 
        \alpha = 1.18^{+0.12}_{-0.06}.
\]
consistent with our previous result. 

A power law representation of the EMSS cluster $\log N-\log S$
distribution has also been computed by Rosati et al.\ (1995) by
integrating the X-ray luminosity functions of Henry et al.\ (1992).
However, Rosati and co-workers do not quote the result of this
exercise in their paper, and their Fig.~5 shows the EMSS power law not
in the EINSTEIN passband but rather in the ROSAT $0.5-2.0$ keV range.
Thus we can only state that the above parameterization is in
qualitative agreement with their figure.

We emphasize once more that the discrepancy between the BCS and EMSS
power law slopes does not imply a conflict between the two samples, in
particular since the actual {\em data}\/ agree at fluxes greater than
$2\times10^{-12}$ erg cm$^{-2}$ s$^{-1}$ ($0.3-3.5$ keV).  Given the
impressive redshift range spanned by the EMSS cluster sample
($0.046\leq z \leq 0.826$), its $\log N-\log S$ distribution can not
be expected to follow a single power law.  Indeed, a trend for the
EMSS $\log N-\log S$ distribution to flatten with decreasing flux is
evident in Fig.~\ref{lognlogs_more}.

Table~\ref{lognlogs_tab} summarizes the fit results for the BCS $\log
N-\log S$ distributions in all five energy bands.

\begin{table}
 \begin{tabular}{l@{\hspace{1mm}}c@{\hspace{2mm}}c@{\hspace{1mm}}c}
     energy range    &       $S_{\rm min}$ range           & 
              $\kappa$ & $\alpha$ \\ 
  \hspace*{3mm}(keV) & ($10^{-11}$ erg cm$^{-2}$ s$^{-1}$) & 
             (sr$^{-1}$ ($10^{-11}$ erg cm$^{-2}$ s$^{-1}$)$^\alpha$) & \\ \hline  \\
  $0.1-\;\;2.4$  & $0.40-0.80$  & $   18.63^{-0.08}_{+0.09}$ & $1.31^{+0.06}_{-0.03}$ \\
  $2.0-10.0   $  & $0.35-0.70$  & $   15.67^{-0.10}_{+0.11}$ & $1.28^{+0.05}_{-0.03}$ \\
  $0.5-\;\;2.0$  & $0.25-0.50$  & $\;\;9.80^{-0.25}_{+0.09}$ & $1.32^{+0.05}_{-0.04}$ \\
  $0.3-\;\;3.5$  & $0.42-0.84$  & $   20.05^{-0.24}_{+0.27}$ & $1.30^{+0.05}_{-0.04}$ \\
  bolometric     & $0.80-1.60$  & $ 47.0^{-1.6}_{+3.5}\;\;\;$ & $1.27^{+0.08}_{-0.05}$ \\ 
 \end{tabular}
 \caption[]{The results of maximum likelihood fits of a power law
          $N(>S) = \kappa \, S^{-\alpha}$ to the BCS $\log N-\log S$
          distributions in five different energy bands. The quoted
          best fit parameters are the median values of the results
          obtained in fits to the data at $S>S_{\rm min}$ where
          $S_{\rm min}$ varied as indicated. The quoted errors are the
          10$^{\rm th}$ and 90$^{\rm th}$ percentiles, respectively.}
          \label{lognlogs_tab}
\end{table}

\subsection{The flux-limited ROSAT Brightest Cluster Sample}

As discussed in detail in Section~\ref{serdet}, the BCS cluster list
starts to become increasingly incomplete at fluxes of about $5 \times
10^{-12}$ erg cm$^{-2}$ s$^{-1}$ because serendipitous VTP detections
that could compensate for the incompleteness of the sample based on
SASS detections alone are available for only about one sixth of the
total BCS sky area. Although this incompleteness can be corrected for
and is thus of no major concern for the study of certain cluster
statistics such as the $\log N-\log S$ distribution (see
Section~\ref{logNlogS}) or the X-ray luminosity function, it
represents a severe handicap for other investigations, most notably
the spatial cluster-cluster correlation function, where such
corrections can not be applied and a reasonably complete
statistical sample is required.

\begin{figure}
  \epsfxsize=0.5\textwidth
  \hspace*{0cm} \centerline{\epsffile{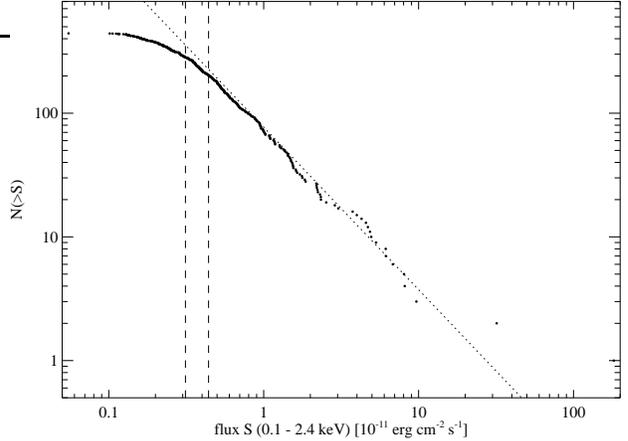}}
  \caption[]{The cumulative flux distribution of all 442 clusters in our
           sample of VTP detections with raw count rates greater than
           0.07 count s$^{-1}$. The dotted line represents the power
           law description of the BCS $\log N-\log S$ distribution as
           determined in Section~\protect\ref{logNlogS}. The dashed
           lines mark the flux limits of 80 and 90 per cent
           completeness, respectively.}  \label{flux_lim}
\end{figure}

Fig.~\ref{flux_lim} shows again the cumulative number counts of all
442 BCS clusters as a function of flux. However, this time all
clusters are given the same weight of unity, i.e., no corrections are
made for variations in the effective sky coverage within our
sample. The overlaid power law represents the best fit to the BCS
$\log N-\log S$ distribution as determined in Section~\ref{logNlogS}.
The requirement that the flux completeness be at least 80 per cent
leads to a limiting flux of $3.1 \times 10^{-12}$ erg cm$^{-2}$
s$^{-1}$ and yields a subsample of 283 clusters. Included in this
flux-limited, 80 per cent complete BCS is a smaller, 90 per cent flux
complete sample of 203 clusters with fluxes higher than $4.4\times
10^{-12}$ erg cm$^{-2}$ s$^{-1}$. Limiting the BCS to this more
complete subsample will allow us to quantify how incompleteness
affects the BCS cluster-cluster correlation function (Edge et al., in
preparation).

\section{The spatial distribution of BCS clusters}

In this section we discuss the redshift and Galactic latitude
distribution of the clusters in the flux limited BCS and present the
BCS clusters' distribution on the sky.

\subsection{The redshift distribution of the BCS}

\begin{figure} 
  \epsfxsize=0.5\textwidth
  \hspace{0cm} \centerline{\epsffile{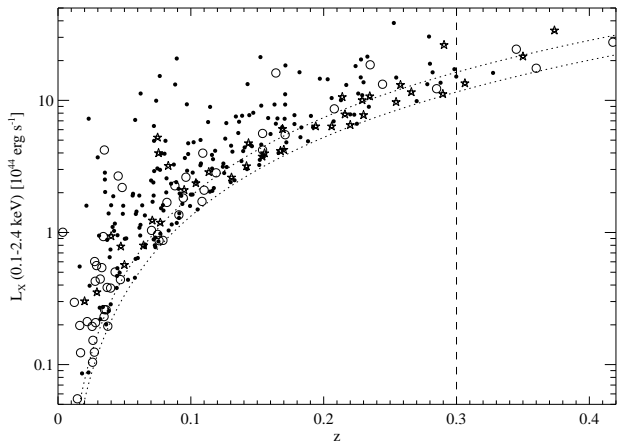}}
  \caption{The X-ray luminosity of the 283 clusters of the 80 per cent
	    complete flux limited BCS as a function of redshift. The
	    dotted lines show the cutoff introduced by the X-ray flux
	    limits at $3.1 \times 10^{-12}$ erg cm$^{-2}$ s$^{-1}$ (80
	    per cent completeness) and $4.4 \times 10^{-12}$ erg
	    cm$^{-2}$ s$^{-1}$ (90 per cent completeness).  Abell
	    (Zwicky) clusters are plotted as solid dots (stars); the
	    remaining 52 clusters not contained in these largest
	    optical cluster catalogues are shown as open circles.}
	    \label{bcs_lum_z}
\end{figure}

Fig.~\ref{bcs_lum_z} shows the BCS clusters' X-ray luminosity as a
function of redshift. (The two components of the double cluster A1758
have been merged for this plot\footnote{The only other double cluster
remaining in the flux limited sample is A2572a($=$HCG94)/A2572b. This
system is at a redshift of 0.04 and, contrary to A1758 ($z=0.28$),
well resolved into two sources at the angular resolution of the
RASS.}.) Of the initial 95 serendipitous VTP-detected clusters (see
Section~\ref{serdet}), 19 are sufficiently X-ray bright to be included
in the 80 percent complete BCS, and for 7 the X-ray flux exceeds also
the flux limit of the 90 percent complete BCS.

We note that although 68 (70) per cent of the clusters in the 80 (90)
per cent complete BCS subsamples are Abell clusters, the fraction of
non-Abell clusters in the BCS is substantial: 13 (11) per cent of the
systems contained in the 80 (90) per cent complete sample are Zwicky
clusters without Abell identification and another 18 (19) per cent are
listed in neither of the two largest optical cluster catalogues used
in our study.

Redshifts have been measured for 98 per cent of the clusters in the 80
per cent complete sample; for the 90 per cent complete sample the
redshift completeness rises to 100 per cent. However, a number of BCS
clusters are (by RASS standards) very distant ($z\ga 0.35$) and fall
thus into a regime where a reliable optical confirmation is no longer
feasible from the POSS images alone. Since the POSS plates were our
main source of optical information in the cluster verification and
confirmation stage of the compilation of the BCS
(cf. Section~\ref{clean_up}), this introduces a redshift limit to the
statistically useful BCS at $z\sim 0.3$ (see the dashed line in
Fig.~\ref{bcs_lum_z}. Within $z=0.3$, 98 (100) per cent of the 80
(90) per cent complete BCS have measured redshifts. The sizes of
these, our final statistical BCS subsamples within $z=0.3$, are 276
and 201 clusters for 80 and 90 percent flux completeness,
respectively\footnote{Subclusters are counted separately: A\,2572a/b
account for two clusters in either sample.}.

\begin{figure} 
  \epsfxsize=0.5\textwidth
  \hspace*{0cm} \centerline{\epsffile{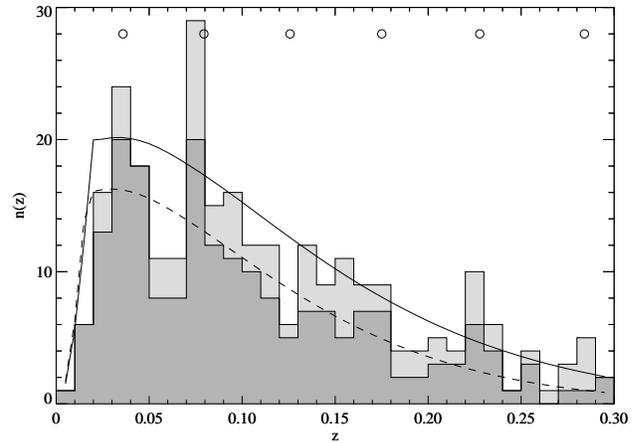}}
  \caption{The differential redshift distribution of the clusters in the
            flux limited BCS. The distribution shown in light shading
            represents the 80 percent complete sample; the 90 percent
            complete subsample is shown in dark shading.  The open
            circles in the upper half of the plot mark the loci of
            spherical shells in comoving space at a constant
            separation of $120\,h_0^{-1}$ Mpc. The solid and dashed
            curves show the redshift distribution expected for a
            spatially homogeneous cluster distribution.}
            \label{bcs_n_z}
\end{figure}

\begin{figure*}
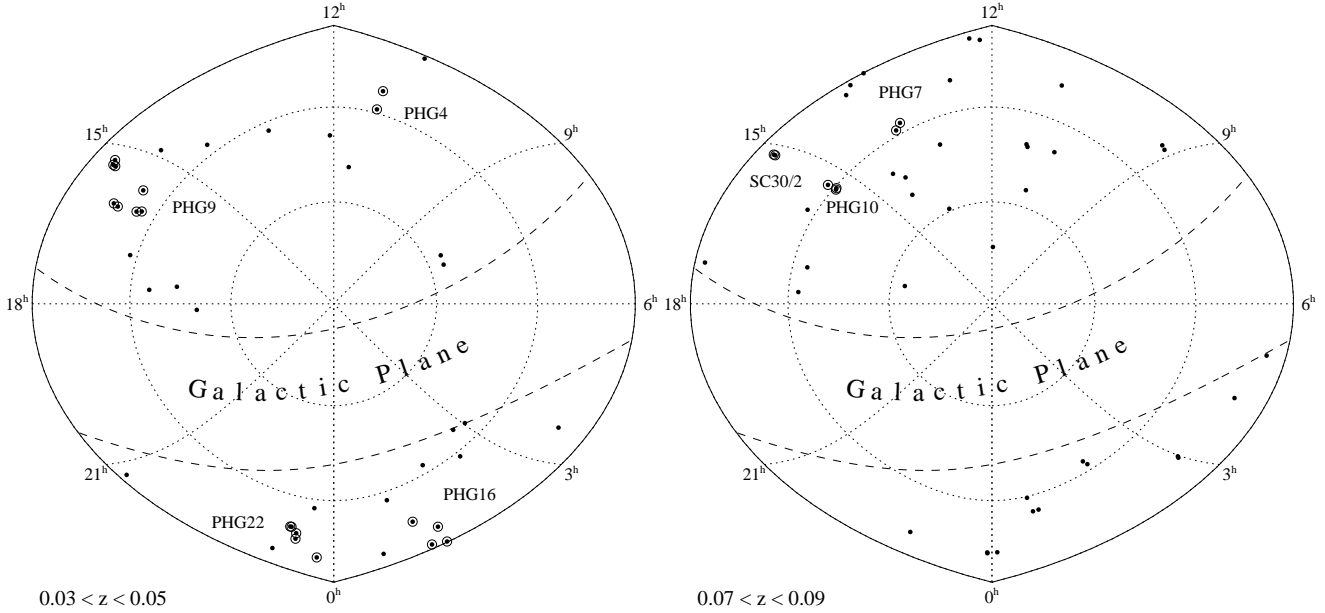
 
  \parbox{0.5\textwidth}{
  \epsfxsize=0.49\textwidth
  \epsffile{bcs_skymap_z1.epsf}}
  \parbox{0.5\textwidth}{
  \epsfxsize=0.49\textwidth
  \epsffile{bcs_skymap_z2.epsf}}
  \caption{The sky distribution of the BCS clusters of the 80 per cent
      	   complete sample that fall into the $0.03 \leq z < 0.05$
      	   (left) and $0.07 \leq z < 0.09$ redshift intervals
      	   (right). Clusters belonging to known Abell superclusters
      	   are plotted encircled. In the case of the Hercules
      	   supercluster (PHG9), we assume membership also for AWM\,4
      	   and MKW\,3s.  An equal-area Aitoff projection is used. The
      	   40 deg wide exclusion zone around the Galactic equator is
      	   marked by the dashed lines.}  \label{bcs_z_skymap}
\end{figure*}

Fig.~\ref{bcs_n_z} shows the redshift distributions of the 80 and 90
per cent complete BCS. Note the two pronounced peaks at redshifts
around 0.035 and 0.08. These are in part due to a series of
superclusters, as can be seen from Fig.~\ref{bcs_z_skymap} which shows
the sky positions of the clusters at the redshifts of these two peaks
in equatorial coordinates. No less than 4 superclusters from the
catalogue of Postman, Huchra \& Geller (1992) contribute to the first
of the peaks in Fig.~\ref{bcs_n_z}. The Hercules supercluster, PHG9,
contributes 6 Abell clusters to the BCS in the redshift range $0.03
\leq z < 0.05$ (A\,2052, A\,2063, A\,2107, A\,2147, A\,2148, A\,2151);
the non-Abell clusters AWM\,4 and MKW\,3s are probably also associated
with this supercluster. Another 5 BCS clusters in the same redshift
range (A\,2572a/b, A2589, A2593, and A2657) are part of the central
region (PHG22) of the Perseus-Pegasus filament. Finally, PHG4 and
PHG16 are represented by the systems A\,1177 and A1185, and A\,147,
A\,160, A\,168, and A\,193, respectively.  Around the redshift of the
second peak in Fig.~\ref{bcs_n_z}, i.e.\ at $0.07 \leq z < 0.09$, it
is mainly the Corona Borealis supercluster (PHG7 and PHG10) that
contributes (A\,1775 and A\,1800, and A\,2061, A\,2065 and A\,2067,
respectively). Additional signal comes from less prominent
superclusters like SC30/2 (Zucca et al.\ 1993) two of whose members
(namely A\,2029, A\,2033) are in this redshift slice of the BCS, as
well as cluster pairs such as A\,399/A\,401 and A\,272/A\,278.

It is clear from Fig.~\ref{bcs_z_skymap} that neither of the
pronounced two first peaks in Fig.~\ref{bcs_n_z} can be attributed to
any single supercluster, although the Hercules supercluster accounts
for about half of the excess signal around $z=0.04$. Rather, it is
clusters and superclusters from all over the sky that create the
observed signature.

Motivated by this result, we investigate whether there is evidence for
further peaks in the BCS redshift distribution.  The spacing of the
peaks at $z\sim 0.04$ and 0.08 corresponds to a separation in comoving
distances of about 120 $h_0^{-1}$ Mpc, a value very close to the
apparent periodicity of 128 $h_0^{-1}$ Mpc found in the galaxy `pencil
beam' surveys of Broadhurst et al. (1990).

If we assume that this is a typical separation intrinsic to
large-scale clustering, we expect further peaks in the BCS redshift
distribution at the locations indicated by the open circles at the top
of Fig.~\ref{bcs_n_z}. Although neither of these more distant peaks is
observed with a significance comparable to that of the first
two\footnote{Note that the apparent peak at $z\sim 0.225$ in
Fig.~\ref{bcs_n_z} (lightly shaded histogram) contains one cluster
with an estimated (rather than measured) redshift of $z\approx 0.23$.
Estimated redshifts do not, however, contribute to any of the other
peaks.}, the overall appearance of the BCS redshift distribution still
lends mild support the hypothesis of some regularity out to higher
redshifts. We stress that this is the first time that evidence for a
semi-regular redshift pattern has been found in a large-scale X-ray
selected cluster sample. As the BCS covers the whole extragalactic
region of the northern hemisphere (i.e., a solid angle of more than
13,000 deg$^2$) and extends out to redshifts of $z=0.3$, any such
regularity can certainly not be attributed to unfortunate sampling as
was argued in the case of the galaxy pencil beam surveys (van de
Weygaert 1991). Our findings are in qualitative agreement with the
explanation of Bahcall (1991) that a whole series of superclusters
contributes to the periodicity found in galaxy pencil beam surveys,
but are in conflict with her prediction that the apparent periodicity
should be washed out when averaged over a large solid
angle. Apparently the tails of superclusters are interconnected over
even larger scales than was previously thought. Our finding thus lends
support to the concept of a characteristic scale of $\sim 115$ Mpc in
the large-scale distribution of matter advocated by, e.g., Einasto et
al.\ (1997) based on optically selected cluster samples.

A more detailed discussion of the large-scale structure traced by the
BCS will be presented in a forthcoming paper.

\subsection{The Galactic latitude distribution of the BCS}

\begin{figure} 
  \epsfxsize=0.5\textwidth
  \hspace*{0cm} \centerline{\epsffile{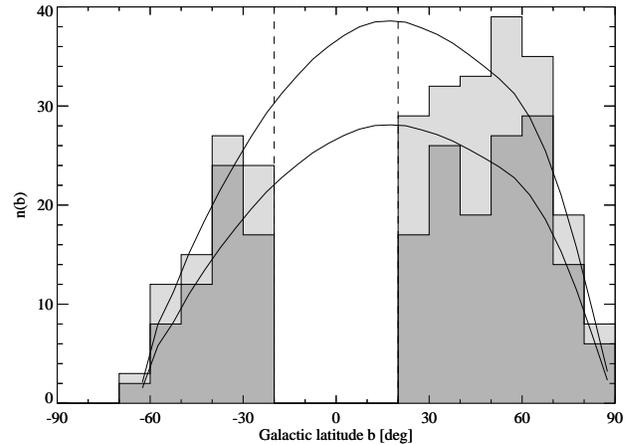}}
  \caption{The differential Galactic latitude distribution of the 276
	   (201) BCS clusters within $z=0.3$ in the 80 (90) percent
	   complete samples (light and dark shaded histograms,
	   respectively). The solid lines indicate the run expected
	   for either sample for a uniform distribution of clusters on
	   the sky.}  \label{bcs_n_gb}
\end{figure}

Fig.~\ref{bcs_n_gb} shows the distribution of the BCS clusters with
Galactic latitude $b$ for the 80 and 90 percent complete samples and
$z\leq 0.3$.  Note that, for both samples, the number of BCS clusters
as a function of $b$ is in agreement with that expected for a uniform
distribution of clusters on the sky (shown as the solid lines in
Fig.~\ref{bcs_n_gb}). There is a slight deficiency in BCS clusters
just north of the Galactic plane; for the 90 percent complete BCS the
dearth of clusters in this region is significant at the $2.1\sigma$
level.  The excess of clusters at Galactic latitudes of about
60$^{\circ}$ is, again, only marginally significant ($<2.1\sigma$) for
either sample. It is not caused by any known supercluster as, of the
PHG superclusters (Postman et al.\ 1992), only PHG4 contributes two
clusters to the BCS at $60^{\circ}\leq b < 70^{\circ}$.

\subsection{The distribution of the BCS clusters in the sky}

\begin{figure*}
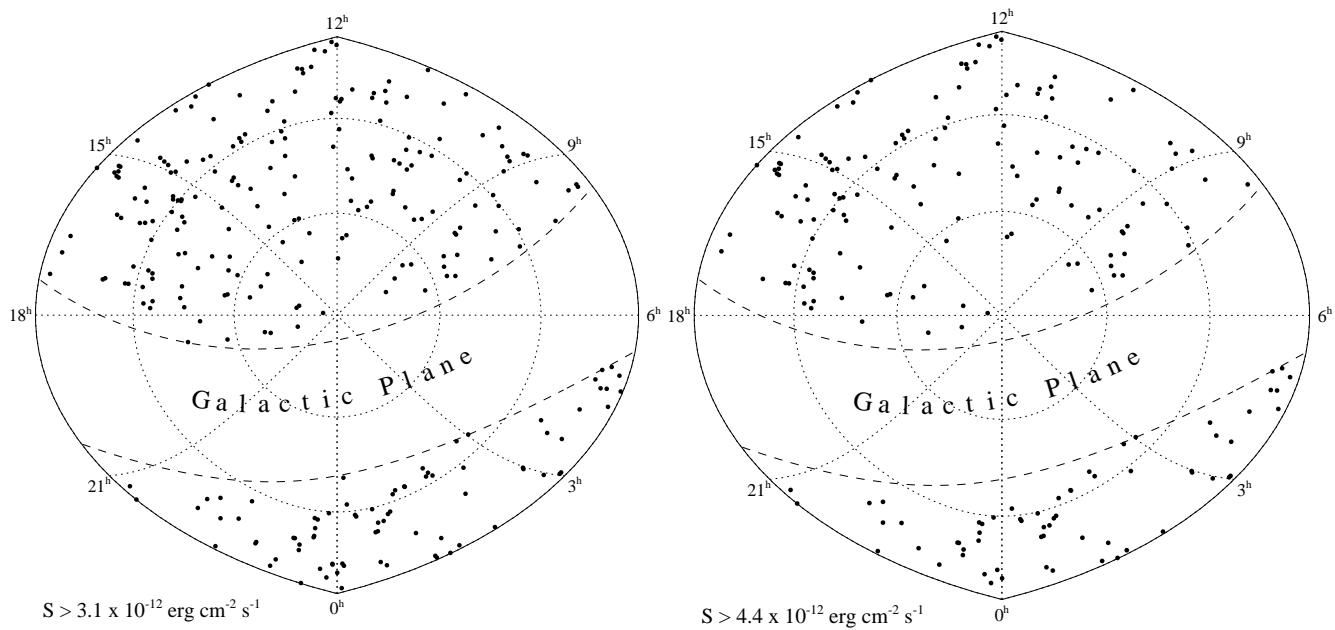
 
  \parbox{0.5\textwidth}{
  \epsfxsize=0.49\textwidth
  \epsffile{bcs_skymap_1.epsf}}
  \parbox{0.49\textwidth}{
  \epsfxsize=0.5\textwidth
  \epsffile{bcs_skymap_2.epsf}}
  \caption{The distribution of the BCS clusters in the 80 and 90 per
           cent sample (left and right panel, respectively) in an
           equal-area Aitoff projection of the northern
           hemisphere. The 40 deg wide exclusion zone around the
           Galactic equator is marked by the dashed lines.}
           \label{bcs_skymap}
\end{figure*}

Fig.~\ref{bcs_skymap} shows the equatorial sky distribution of the
clusters in the two flux-limited BCS subsamples. Again, a redshift
cutoff at $z=0.3$ has been applied.

\section{Differences between the Abell content of the BCS and the XBACs}

Although the XBACs sample and the BCS have been compiled from the same
RASS data, the Abell subsample of the flux limited BCS does not
exactly match the XBACs list where both samples overlap in sky area
and X-ray flux. The small, existing differences are due to different
classifications for a few clusters and small variations in the final
fluxes caused by, firstly, the use of more accurate flux correction
factors in the BCS (Section~\ref{flux_corr}), as well as, secondly,
somewhat different criteria concerning the application of these
correction factors for pointlike and extended sources (see
Section~\ref{flux_corr} of this paper and Section~6.1 in EVB).

In detail the differences are as follows: The XBACs sample contains
three clusters missing from the subsample of the BCS obtained by
applying the XBACs flux limit of $5.0 \times 10^{-12}$ erg cm$^{-2}$
s$^{-1}$, namely A\,1035, A\,2228, and A\,2637. A\,1035 is listed as a
cluster in both samples but its BCS flux is considerably lower than
the XBACs value due to the fact that a nearby soft X-ray source that
was considered part of the cluster emission during the compilation of
the XBACs sample was, upon careful reassessment of the optical and
X-ray evidence, discarded as unrelated in the course of the BCS
compilation. A\,2228 is treated as a cluster with severe AGN
contamination in the XBACs list, whereas it was discarded from the BCS
as being dominated by AGN contribution. A\,2637 (Ledlow \& Owen 1996)
is a similar case: the system was included in the XBACs sample and
marked as contaminated by emission from a background AGN, whereas the
same cluster is missing from the BCS as the X-ray emission is now
classified as being dominated by the same AGN.

For ten Abell clusters the opposite scenario is realized: they are
contained in the BCS and listed as featuring fluxes of, or greater
than, $5.0 \times 10^{-12}$ erg cm$^{-2}$ s$^{-1}$ but are nonetheless
missing from the XBACs sample. For two of these there has been a
change of emphasis in the opposite sense to that mentioned above: the
BCS lists them as {\em contaminated}\/ by AGN emission while they were
discarded from the XBACs sample as likely to be {\em dominated}\/ by
AGN emission.

The fact that only three Abell clusters are contained in the XBACs
sample but not in the corresponding subset of the BCS, while the
opposite is true for ten clusters, is due to the mentioned small
changes in the final fluxes: the median ratio of the BCS and XBACs
fluxes for clusters contained in both samples is 1.015, i.e. there is
a tendency for the more accurate BCS fluxes to be slightly higher than
the XBACs fluxes computed earlier. This causes eight Abell clusters to
just make the XBACs flux limit of $5.0 \times 10^{-12}$ erg cm$^{-2}$
s$^{-1}$ when the BCS fluxes are used, while they remain just below
the flux limit in the original XBACs list.

For all practical purposes the sample of Abell clusters with fluxes
higher than $5\times 10^{-12}$ erg cm$^{-2}$ s$^{-2}$ in the BCS can
thus be considered to be the same as the subset of the XBACs sample
defined by $|b| \ge 20^{\circ}$ and $\delta \ge 0^{\circ}$.

\section{The relevance of the 1RXS catalogue for the BCS}
\label{rassbsc}

While this paper was prepared for submission, MPE released the ROSAT
All Sky Survey Bright Source Catalogue (1RXS, Voges et al.\ 1996), a
list of more than 18,000 sources down to count rates of 0.05 count
s$^{-1}$ detected by the SASS in its second processing of the RASS
data. The main difference between the first SASS processing (SASS-I)
that the compilation of the BCS started from (see
Section~\ref{rass_db}) and this second processing (SASS-II) is that
the analysis is no longer confined to 2 degree wide strips but is
performed on merged photon maps of the same type as those used by VTP
in our re-processing of the RASS data. Note, however, that while the
raw data used by the SASS have thus changed, the SASS detection
algorithms have not.  Accordingly, the severe underestimation of the
flux from extended emission discussed by EVB persists and a
re-processing of the raw data (by VTP or another suitable algorithm)
remains mandatory for the compilation of an X-ray flux limited cluster
sample.

Besides the differences in the underlying raw data, the source lists
from the first SASS processing and the 1RXS catalogue differ also in
the applied source likelihood threshold. Whereas the SASS-I master
source list contains all detections with a likelihood of existence
higher than 10, a more conservative threshold value of 15 was used in
the compilation of the 1RXS catalogue\footnote{The version released by
MPE on 19 June 1996 also contains 58 sources with count rates and
existence likelihoods of zero. In the latest release (1.2), this
number has dropped to 9.}. Probably due to this higher threshold, the
1RXS catalogue lists only 8,541 sources with count rates higher than
0.1 count s$^{-1}$ as compared to 10,241 found in the first SASS
processing (see Section~\ref{sass_sample}).

Given that the number of 1RXS sources above our initial count rate cut
is smaller than in SASS-I we do not necessarily expect a 1RXS based
cluster sample to be more complete than the BCS as it stands. To test
this explicitly, we repeated the first steps of the BCS compilation
procedure using the 1RXS catalogue (limited to our study area and
sources brighter than 0.1 count s$^{-1}$) as the input X-ray source
list. We find the number of true coincidences in the
cross-correlations with the Abell cluster catalogue to lie 10 per cent
below the figure obtained for the source list from the first SASS
processing while the number of true coincidences with Zwicky clusters
is essentially the same as for the SASS-I source list (see
Section~\ref{sass_sample}). The number of significantly extended
sources in the 1RXS catalogue (318 at $\delta \geq 0^{\circ}$, $|b|
\geq 20^{\circ}$ and with count rates greater than 0.1 count s$^{-1}$)
is significantly lower than the one obtained in the first SASS
processing (406 within the same constraints).

A cluster sample compiled from the 1RXS catalogue following the BCS
procedure would thus probably be smaller, i.e.\ less complete, than
the BCS as presented here. If we consider only those clusters in the
final 90 per cent flux complete BCS subsample that have SASS-I count rates
higher than 0.1 count s$^{-1}$ (i.e.\ excluding serendipitous VTP
detections), we find that 10 of the 196 SASS-I-detected BCS clusters
are not contained in the 1RXS catalogue {\em at any count rate limit}.
Among these ten is Virgo, the X-ray brightest cluster in the sky, as
well as three other clusters with VTP fluxes in excess of $1\times
10^{-11}$ erg cm$^{-2}$ s$^{-1}$ (A\,2589, A\,2593, and A\,76). If we
apply the same count rate cut to the 1RXS catalogue as before to the
SASS-I source list (0.1 count s$^{-1}$), the number of BCS clusters
missing from the 1RXS list becomes even higher: no fewer than 23 of the
201 X-ray brightest clusters in the northern hemisphere would be
missing had we used the 1RXS catalogue instead of the SASS-I source
list\footnote{We consider a BCS cluster as missing from the 1RXS
catalogue if the latter contains no entry within 10 arcmin (radius) of
the VTP cluster position.}. For the seven clusters in the 90 per cent
complete BCS that were serendipitously detected by VTP the situation
is even worse: two have been missed altogether by SASS-II, another one
is listed with a count rate below our threshold of 0.1 count s$^{-1}$,
and a fourth one, RXJ1206.6+2811 (MKW4s), lies above the count rate
limit but is classified as a point source and thus would also not have
made it into the BCS.

As detailed above, more than 10 per cent of the clusters in the 90 per
cent complete BCS would be missing had we based our compilation on the
SASS-II detections listed in the 1RXS catalogue rather than on the
SASS-I master source list. Although we cannot rigorously prove that
the fact that SASS-II fails to reproduce the 90 per cent complete BCS
is a reliable measure of the overall performance of the second
processing, we are confident that the number of clusters that have
been detected by SASS-II outside the area re-processed by VTP, and
which may turn out to be bright enough to be included in the BCS when
re-processed by VTP, cannot outweigh this incompleteness.

\section{BCS -- the sample}

Table~\ref{bcs_tab} lists the 201 X-ray brightest clusters with
Galactic latitudes $|b| \geq 20^{\circ}$, within a redshift of 0.3,
and with 0.1 to 2.4 keV fluxes greater than $4.4 \times 10^{-12}$ erg
cm$^{-2}$ s$^{-1}$. This sample is 90 percent flux complete. We also
include 2 clusters at redshifts greater than 0.3 (marked with a
$\star$ symbol in the first column) that otherwise meet the selection
criteria. For completeness' sake we also list A\,2151b, an apparent
low-luminosity satellite of A\,2151a, although its flux falls almost a
factor of three short of the quoted limit, as well as A\,1758a/b which
are detected as separate subclusters below the flux limit but are
bright enough to make the flux cut when combined.  Of these 201
clusters, 71 per cent are Abell clusters, 10 per cent are Zwicky
clusters, and 19 per cent are systems not contained in either of these
two largest optical cluster catalogues. The diversity of our final
sample underlines yet again the importance of X-ray selected samples
for unbiased statistical studies of cluster properties.

The sample listed explicitly in Table~\ref{bcs_tab} is limited to a
size of $\sim 200$ clusters in order to keep it smaller than a cluster
sample under compilation in a second follow-up project of northern
RASS clusters (B\"ohringer et al., in preparation, Burg et al., in
preparation). An extended BCS listing including clusters down to a
flux limit of $\sim 3 \times 10^{-12}$ erg cm$^{-2}$ s$^{-1}$ will be
published later.

In detail the contents of Table~\ref{bcs_tab} are
\begin{list}{}{\labelwidth17mm \leftmargin17mm}
  \item[column\,\,\, 1:] redshift, contamination, extent, and serendipity flag.
		   Clusters at $z>0.3$ are marked $\star$; `c' means a 
		   significant fraction of the quoted flux may come from 
                   embedded point sources; `V' (`S') signals significant X-ray
                   extent according to VTP (SASS), i.e., an extent value in 
                   excess of 35 arcsec; systems flagged by a $\bullet$ symbol 
                   are serendipitous VTP detections in the sense of 
 		   Section~\ref{serdet}.
  \item[column\,\,\, 2:] cluster name. Where clusters appear to consist of two 
		   components, two entries (`a' and `b') are listed.
  \item[column\,\,\, 3:] Right Ascension (J2000) of the X-ray position as
                   determined by VTP.
  \item[column\,\,\, 4:] Declination (J2000) of the X-ray position as 
                   determined by VTP.
  \item[column\,\,\, 5:] Column density of Galactic Hydrogen from Stark et al.
                   (1992).
  \item[column\,\,\, 6:] RASS exposure time (accumulated).
  \item[column\,\,\, 7:] PSPC count rate in PHA channels 11 to 235 originally
                   detected by VTP.
  \item[column\,\,\, 8:] The equivalent radius $\sqrt{A_{\rm VTP}/\pi}$ of
                   the source detected by VTP. 
  \item[column\,\,\, 9:] Final PSPC count rate in Pulse Height Analyzer (PHA) 
                   channels 11 to 235
                   based on the original VTP count rate. Statistical 
		   corrections for low-surface
                   brightness emission that has not been detected directly and
                   for contamination from point sources have been applied.
  \item[column 10:] Error in the final PSPC count rate according to 
		   equation~\ref{vtp_err}. The fractional uncertainty in the 
		   energy flux (column 13) and the X-ray luminosity (column 14)
 		   can be assumed to be the same as the fractional count rate 
		   error.
  \item[column 11:] ICM gas temperature used in the conversion from count
 		   rates to energy fluxes. `e' indicates the temperature has
		   been estimated from the $L_{\rm X}-{\rm k}T$ relation.
  \item[column 12:] Redshift. 
  \item[column 13:] Unabsorbed X-ray energy flux in the 0.1 to 2.4 keV band.
  \item[column 14:] Intrinsic X-ray luminosity in the 0.1 to 2.4 keV band
		   (cluster rest frame).
  \item[column 15:] Reference for the redshift in column 12.
\end{list}

The source positions listed in columns 3 and 4 of Table~\ref{bcs_tab}
are typically good to about 1 arcmin. Depending on the angular extent
of the emission (as indicated by the equivalent detection radius
quoted in column 8) and the presence of substructure and/or
contaminating point sources (usually unknown) somewhat larger
positional errors for individual sources can not be ruled out.

\begin{table}
 \caption{The ROSAT Brightest Cluster Sample (see appended pages)}
 \label{bcs_tab}
\end{table}

\noindent

\section{Summary} 

\begin{figure*} 
  \epsfxsize=19cm
   \vspace*{0cm}
  \epsffile{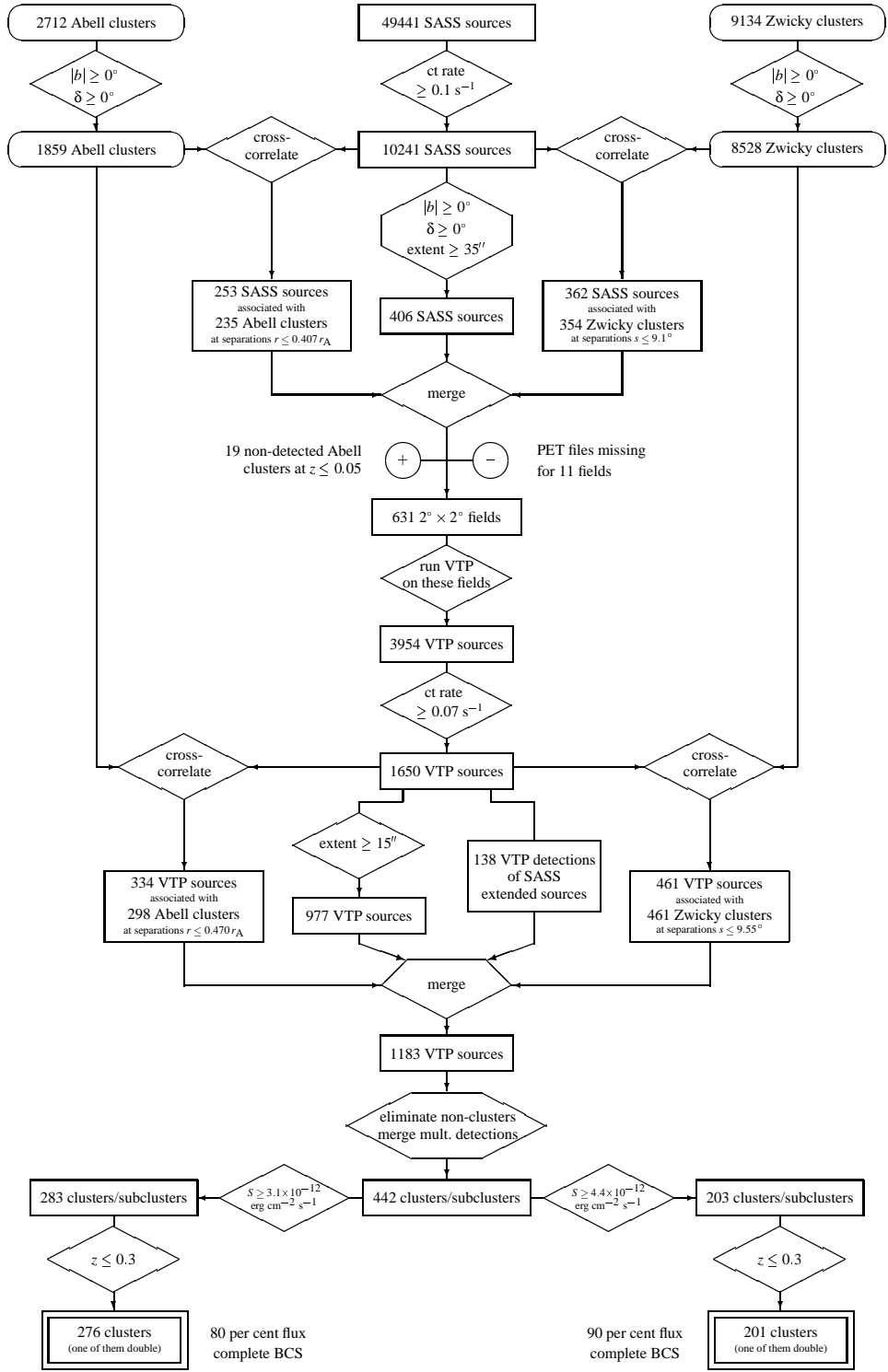} \mbox{}\\*[-5cm]
  \caption{Flow diagram summarizing the compilation of the BCS.}
  \label{bcs_flow}
\end{figure*}

We present the ROSAT Brightest Cluster Sample (BCS), an X-ray flux
limited sample of clusters of galaxies in the northern extragalactic
sky ($\delta \geq 0^{\circ}$, $|b| \geq 20^{\circ}$). Although the
bulk of the BCS clusters are Abell clusters ($\sim 70$ per cent), the
sample contains also a considerable number of Zwicky clusters ($\sim
10$ per cent) as well as purely X-ray selected systems ($\sim 20$ per
cent). In terms of being a representative cluster sample the BCS
constitutes a significant improvement over the all-sky XBACs sample of
EVB which is of similar size but limited to ACO clusters.  The
procedures followed during the compilation of the BCS are in many ways
very similar to that devised by EVB for the compilation of the XBACs
sample. An overview of the compilation procedure is given in
Fig.~\ref{bcs_flow}.

We confirm the finding of EVB that the compilation of a properly flux
limited cluster sample from the RASS data is impossible from SASS
output alone. The re-processing of the RASS photon event files with
VTP, a source detection algorithm developed specifically for the
detection of low-surface brightness emission of essentially arbitrary
geometry, is thus of vital importance for the compilation of the BCS.
A comparison with the VTP count rates for clusters with $S> 2\times
10^{-12}$ erg cm$^{-2}$ s$^{-1}$ shows the SASS count rates to be too
low by typically a factor of 0.64 (median) with the 10$^{\rm th}$ and
90$^{\rm th}$ percentiles being 0.35 and 0.84. Any SASS count rate
limited cluster sample will thus necessarily be incomplete. Since the
compilation of the BCS started from just that, i.e.\ a count rate
limited SASS source list, and VTP was run only on part of our study
area, our final cluster sample is incomplete even at fluxes as high as
$4 \times 10^{-12}$ erg cm$^{-2}$ s$^{-1}$ ($0.1-2.4$ keV). However, a
set of 95 clusters detected serendipitously by VTP allows this
incompleteness to be accurately quantified and corrected for.

After correction for incompleteness, we find the $\log N - \log S$
distribution of our clusters to follow a power law of slope $-1.31$
down to X-ray fluxes of about $2 \times 10^{-12}$ erg cm$^{-2}$
s$^{-1}$. Although the canonical Euclidean value of $-1.5$ is formally
ruled out by our measurement, the observed flattening of the $\log N -
\log S$ distribution with decreasing flux is not indicative of cluster
evolution.  Rather, it can be attributed to cosmological effects in an
expanding Universe. If these are taken into account we find the BCS
$\log N - \log S$ distribution to be consistent with a non-evolving
cluster population down to the quoted flux level.  Using the
best-fitting power law to quantify the level of completeness of the
BCS cluster list, we compile two flux limited subsamples with limiting
fluxes of $3.1 \times 10^{-12}$ and $4.4 \times 10^{-12}$ erg
cm$^{-2}$ s$^{-1}$. The flux completeness of these samples is 80 and
90 percent, respectively. Since clusters at a redshift higher than
about 0.3 can not be reliably confirmed with the presently available
optical plate material, thus requiring additional optical follow-up,
we limit our statistical sample to systems at $z\leq 0.3$. Within this
redshift, the 80 and 90 per cent complete BCS contain 276 and 201
clusters, respectively, if subclusters are counted separately. All
clusters in the 90 per cent complete sample have measured redshifts as
listed in Table~\ref{bcs_tab}.

The redshift distribution of the BCS shows intriguing evidence for
large scale inhomogeneities in the form of cluster overdensities at
certain redshifts. We stress that the observed peaks in the redshift
distribution can only partly be explained by contributions from any
single supercluster.

We assess the statistical quality of strictly X-ray selected cluster
samples by comparing the X-ray extent and hardness ratio distributions
for clusters and point sources. Cluster samples with a high degree of
completeness (85 to 90 per cent) can be compiled from VTP detections
if thresholds are applied to both the X-ray extent and the spectral
hardness of the selected sources. The point source contamination of
the resulting cluster samples will be about 55 per cent; if an
improved VTP algorithm were used the contamination could be reduced to
some 35 per cent. This is also the level of contamination expected for
a purely X-ray selected cluster sample based on SASS detections.
However, due to the inefficiency of the SASS extent criterion, the
completeness of a SASS-selected cluster sample will always remain
below 70 per cent.

The BCS is the largest X-ray selected, flux limited sample of clusters
of galaxies compiled to date. Besides its obvious importance as a
target list for on-going and future X-ray missions, the BCS will allow
detailed studies of a whole series of cluster statistics, such as the
spatial two-point correlation function or the cluster X-ray luminosity
function (Ebeling et al.\ 1997), to name but two.

\section*{Acknowledgements} 

The authors are indebted to the ROSAT team at MPE for making it all
possible and, in particular, for providing the RASS photon data this
analysis is based upon.  We would like to thank especially Cristina
Rosso for her unfaltering support with the RASS PET file retrieval.

The availability of measured redshifts for as many BCS clusters as
possible is of crucial importance for this project. We thus greatly
appreciate the `BCS support observations' performed by Pat Henry and
Chris Mullis which resulted in redshifts for A\,68, A\,781, and
Z\,1665 (data reduction: HE).  

The identification of non-cluster sources that contaminated the sample
we originally started from was greatly facilitated by the availability
of digitized optical images from the POSS and UK Schmidt sky surveys
obtained through the {\sc SkyView} facility and STScI's DSS Web
interface. Thanks to Thomas McGlynn, Keith Scollick and co-workers for
developing and maintaining {\sc SkyView}. All of the data analysis
presented in this paper was carried out using the Interactive Data
Language (IDL). We are greatly indebted to all who contributed to the
various IDL User's Libraries; the routines of the IDL Astronomy User's
Library (maintained by Wayne Landsman) have been used particularly
extensively.

HE is grateful to Pat Henry for fruitful discussions and persistent
moral support. Isabella Gioia and Gerry Luppino provided valuable
background information about the EMSS cluster sample. We thank the
referee, Bob Nichol, whose extensive criticism lead to changes which
significantly improved this paper.

HE gratefully acknowledges partial financial support from a European
Union EARA Fellowship and SAO contract SV4-64008. ACE, ACF and SWA
thank the Royal Society for support. CSC acknowledges financial
support from a PPARC Advanced Fellowship. HB's work is supported by
BMFT through the Verbundforschung programme.

This research has made use of data obtained through the High Energy
Astrophysics Science Archive Research Center Online Service, provided
by the NASA-Goddard Space Flight Center, the Leicester Database and
Archive Service's XOBSERVER programme, and the NASA/IPAC Extragalactic
Database (NED).

\appendix

\section{The statistical quality of strictly X-ray selected RASS 
         cluster samples}
\label{xraysel}

Having selected our final flux limited sample, we can now assess the
efficiency of cluster selection by X-ray properties alone. To this
end, we compare the distributions of the two relevant source
characteristics, X-ray extent and spectral hardness, for point sources
and clusters.

Since, in the conversion from count rates to energy fluxes, different
spectral models have to be assumed for clusters and pointlike sources
such as stars or AGN, this comparison can not be performed as a
function of source flux. We use the detected VTP count rate instead,
and compute the count rate limits which, for clusters of galaxies,
correspond to limiting fluxes of $2\times 10^{-12}$, $3.1\times
10^{-12}$, and $4.4\times 10^{-12}$ erg cm$^{-2}$ s$^{-1}$, i.e., the
lowest possible flux limit for an X-ray selected cluster sample based
on the RASS data in the northern hemisphere, and the BCS flux limits
of 80 and 90 percent flux completeness.

\subsection{Source extent}
\label{extent}

\begin{figure}
  \epsfxsize=0.5\textwidth
  \hspace*{-6mm}
  \epsffile{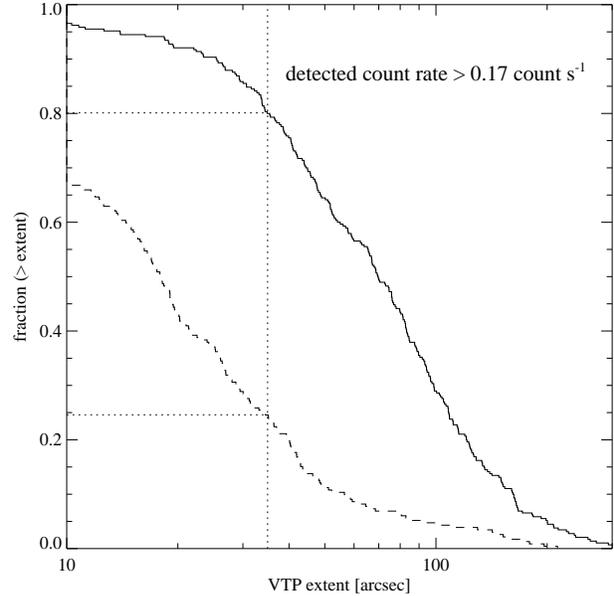}
  \caption[]{The cumulative extent distribution of galaxy clusters
           (solid line) in our sample of VTP detections and the
           corresponding extent distribution of non-cluster sources of
           comparable brightness (dashed line). The count rate cut at
           0.17 count s$^{-1}$ corresponds roughly to the BCS flux
           limit of 80 per cent completeness at $3.1\times 10^{-12}$
           erg cm$^{-2}$ s$^{-1}$.  The dotted lines mark the fraction
           of clusters and non-cluster sources classified as extended
           by VTP when the extent threshold is set to 35 arcsec; the
           intercepts with the ordinate change only slightly when
           count rate cuts corresponding to $2 \times 10^{-12}$ or
           $4.4 \times 10^{-12}$ erg cm$^{-2}$ s$^{-1}$ are chosen.}
           \label{vtp_extent}
\end{figure}

Fig.~\ref{vtp_extent} shows the cumulative, fractional distributions
of the VTP extent parameter for BCS clusters and other, non-cluster
sources at the intermediate of the three quoted count rate limits.
For an extent threshold of 35 arcsec, some 80 per cent of the clusters
are recognized as extended. We find this fraction to change only by a
few per cent when all VTP detections brighter than $2\times 10^{-12}$
erg cm$^{-2}$ s$^{-1}$ are considered or only the ones featuring
fluxes in excess of $4.4\times 10^{-12}$ erg cm$^{-2}$ s$^{-1}$, the
flux limit of the 90 percent complete flux limited BCS. Although a
cluster sample selected by VTP extent alone would thus be highly
complete (note that a completeness of about 90 per cent is achievable
at all flux levels if the extent threshold is lowered to 25 arcsec),
there is a massive penalty for this high degree of completeness: Some
25 per cent of all non-cluster sources, too, are classified as
extended by VTP, again almost independent of the chosen flux limit.
By comparison, the fraction of point sources misclassified as extended
by the SASS is less than 10 per cent at the same numerical extent
value of 35 arcsec (Ebeling et al., 1993).

The reason for the high fraction of point sources erroneously
classified as extended by VTP is twofold. Firstly, there is the effect
of the sources' spectral hardness. The flux correction procedure that
determines the X-ray extents of VTP-detected sources (see
Section~\ref{flux_corr}) uses the PSPC point spread function for a
nominal photon energy of 1 keV. Since the width of the PSF increases
with decreasing photon energy, very soft sources will appear extended
if a too narrow, high-energy PSF is used in the extent determination.
This effect accounts for about 25 per cent of the erroneous
extents. It could, in principle, be heavily suppressed by matching the
PSF to the spectral hardness of each individual source (see
Section~\ref{hardness} for a discussion of the significance of X-ray
hardness ratios).

The second cause for erroneous VTP extents is of a more complex
nature. Recent simulations by Scharf et al.\ (1997, see also Jones et
al.\ 1998) indicate that a large fraction of VTP's erreoneous extent
classifications may be caused by the inclusion of `positive noise' in
the percolation stage of the detection process. This problem can,
however, be almost completely overcome by raising the detection
threshold by some 30 per cent with respect to the surface brightness
of the background.  Although this leads necessarily to lower detected
fluxes, it turns out that the loss in detected flux is entirely
compensated for in the flux correction step described in detail in
Section~\ref{flux_corr}.  According to Scharf and co-workers this
procedure reduces the fraction of point sources erroneously classified
as extended to less than 10 per cent, i.e., a value comparable to that
obtained with the SASS.

\subsection{Spectral hardness}
\label{hardness}

\begin{figure}
  \epsfxsize=0.5\textwidth
  \hspace*{0cm} \centerline{\epsffile{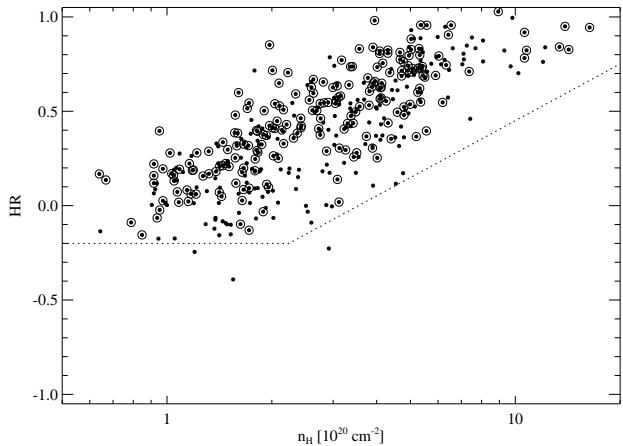}}
  \caption[]{The X-ray spectral hardness ratio as a function of the
	   Galactic column density of neutral hydrogen for all BCS
	   clusters with detected count rates in excess of 0.11 count
	   s$^{-1}$. The dotted line marks the 99 per cent lower limit
	   of the distribution. Note that none of the clusters of the
	   90 percent complete sample (shown encircled) falls below
	   the dotted line.}  \label{bcs_hr}
\end{figure}

The second source characteristic that is a promising discriminator
between clusters of galaxies and other X-ray sources is the spectral
hardness of the detected emission. Following EVB, we define the
hardness ratio HR as
\[ 
          {\rm HR} = \frac{h-s}{h+s},
\]
where $h$ are the photon counts in the hard energy range from 0.5 to 2
keV, and $s$ those in the softer 0.1 to 0.4 keV band. Using the same
definition of HR, Ebeling et al.\ (1993) found more than 90 per cent
of all Abell clusters detected by the SASS in the RASS data to feature
HR values greater than zero. A more stringent criterion can be
obtained by taking into account the X-ray absorbing column density of
Galactic hydrogen, $n_{\rm H}$, in the direction of the cluster (EVB,
Fig.~10).  Fig.~\ref{bcs_hr} shows the observed hardness ratios as
determined by VTP as a function of $n_{\rm H}$ for all BCS clusters
with detected count rates greater than $\sim 0.11$ count s$^{-1}$, the
count rate equivalent of the lowest possible RASS flux limit of
completeness at $2\times 10^{-12}$ erg s$^{-2}$ cm$^{-2}$.  The dotted
line, given by
\begin{equation}
   {\rm HR_{min}} = \max \,(-0.55 + \log \frac{n_{\rm H}}{10^{20} 
                    \;{\rm cm}^{-2}};\,-0.2),
   \label{hrmin}
\end{equation}
marks a lower limit to the hardness ratio of X-ray emission from
groups and clusters of galaxies (99 per cent confidence). It is also a
98 per cent lower limit to the HR values of the VTP detections of ACO
clusters of the extended (unpublished) sample of EVB.

\begin{figure}
  \epsfxsize=0.5\textwidth
  \hspace*{-6mm}
  \epsffile{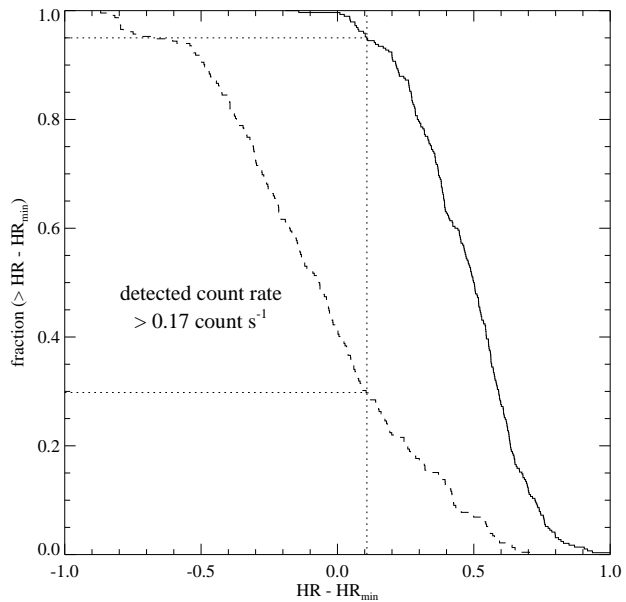}
   \caption[]{The cumulative hardness ratio distribution for the galaxy
           clusters (solid line) in our sample of VTP detections and
           the corresponding HR distribution for non-cluster sources
           of comparable brightness (dashed line). The count rate cut
           0.17 count s$^{-1}$ corresponds roughly to the BCS flux
           limit of 80 per cent completeness at $3.1\times
           10^{-12}$. The dotted lines mark the loci of 95 per cent
           completeness for clusters of galaxies; its intercept with
           the abscissa moves to slightly lower (higher) values when
           count rate cuts corresponding to $2 (4.4) \times 10^{-12}$
           erg cm$^{-2}$ s$^{-1}$ are chosen.}  \label{vtp_hr}
\end{figure}

As can be seen from Fig.~\ref{vtp_hr}, discarding all VTP sources with
spectral hardness ratios below HR$_{\rm min}$ as given by
equation~\ref{hrmin} excludes about 70 per cent of all point sources
while leaving the cluster detections essentially unaffected. Owing to
a weak correlation between spectral hardness and source brightness for
clusters, and a weak anti-correlation between the same quantities for
point sources, the HR threshold resulting in a 95 per cent complete
cluster sample can be increased slightly when only sources above the
count rate limits of our 80 and 90 per cent complete BCS supsamples
are considered.

Although a cluster sample selected by X-ray hardness ratio alone will
thus be highly complete, the remaining contamination from point
sources will still be significant and, in fact, even stronger than for
a purely X-ray extent selected sample.

\subsection{Combining source extent and hardness ratio}

A high fraction of point sources erroneously classified as extended or
a significant fraction of point sources with cluster-like hardness
ratios lead to a severe contamination by point sources for an X-ray
selected cluster sample.  Since, in the RASS data, stars, X-ray
binaries, and AGN together outnumber groups and clusters of galaxies
by about 10 to 1, a total fraction $p$ of hard or extended point
sources will result in a point source contamination of $10p/(10p+c)$
where $c$ is the chosen completeness of the sample. If, for instance,
the contamination by point sources is to be limited to 50 per cent
and, at the same time, 90 per cent completeness are aimed for, then
$p$ is required to be less than 0.09.

It is obvious from Figs.~\ref{vtp_extent} and \ref{vtp_hr} that this
goal cannot be achieved by setting thresholds to either source extent
or hardness ratio alone. However, as these two source properties are
essentially uncorrelated, a combination of both should greatly improve
our odds.

\begin{figure}
  \epsfxsize=0.5\textwidth
  \hspace*{-6mm}
  \epsffile{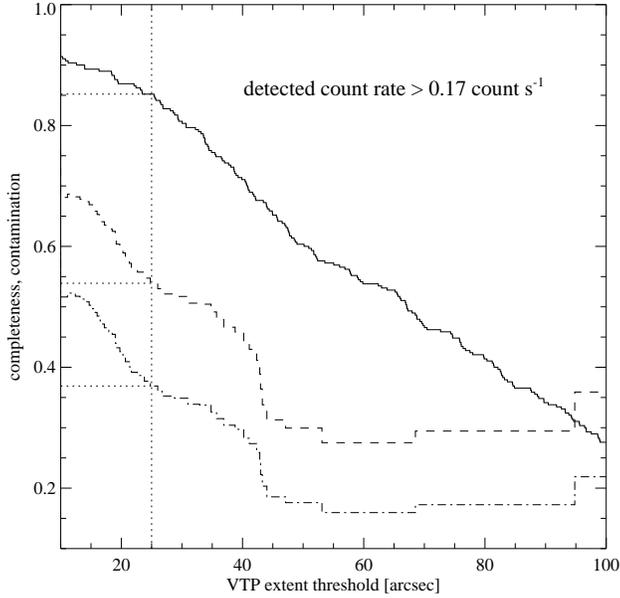}
   \caption[]{Completeness (solid line) and contamination by point
	   sources (dashed line) of a strictly X-ray-selected cluster
	   sample as a function of the chosen VTP extent threshold
	   after the HR cut indicated by the dotted line in
	   Fig.~\protect\ref{vtp_hr} has been applied. The ratio of
	   point sources to extended sources in the overall sample of
	   VTP detections is assumed to be 10 to 1. The count rate cut
	   at 0.17 count s$^{-1}$ corresponds roughly to the flux
	   limit of 80 per cent completeness at $3.1 \times 10^{-12}$
	   erg cm$^{-2}$ s$^{-1}$. The dot-dashed line represents the
	   contamination level achievable with an improved VTP
	   algorithm that suppresses the inclusion of `positive
	   noise'. The dotted lines mark the completeness and
	   contamination levels of a HR-filtered sample when the
	   extent threshold is set to 25 arcsec.}  \label{vtp_xselect}
\end{figure}

Fig.~\ref{vtp_xselect} shows completeness and contamination of an
X-ray selected cluster sample as a function of the chosen extent
threshold if a HR cut as shown in Fig.~\ref{vtp_hr} is applied. We
assume again that the ratio of point sources to intrinsically extended
sources (galaxies as well as groups and clusters of galaxies) in the
RASS is 10 to 1. According to Fig.~\ref{vtp_xselect}, more than 85 per
cent completeness are achievable if a VTP extent threshold of about 25
arcsec is chosen to discriminate between extended and point-like
emission. The contamination of such a cluster sample selected
exclusively by X-ray source characteristics (extent and hardness
ratio) is typically 55 per cent. For an improved VTP algorithm with a
fraction of erroneous extent determinations for point sources half as
high as at present (see Section~\ref{extent}) the contamination is
expected to fall to about 35 per cent (dot-dashed line in
Fig.~\ref{vtp_xselect}). Again, the completeness will be slightly
lower (higher) and the contamination higher (lower) when a lower
(higher) count rate or flux limit is aimed for.

X-ray selected RASS cluster samples of impressive size could be
obtained if such a selection was performed over the whole
extragalactic sky: for flux limits of $2 \times 10^{-12}$, $3.1 \times
10^{-12}$, and $4.4 \times 10^{-12}$ erg cm$^{-2}$ s$^{-1}$ we predict
sample sizes of 1300, 740, and 470 clusters, respectively (100 per
cent completeness).

\subsection{SASS vs VTP: a performance comparison}
\label{sassvsvtp}

As stated in Section~\ref{extent}, the SASS is a priori less prone to
misclassifications of point sources as extended than VTP, while the
SASS hardness ratio is as reliable a discriminator as is the one
derived by VTP. As a consequence, one might expect that cluster
selection by X-ray properties should be even more efficient when the
source parameters returned by the SASS are used. This, however, is
true only as far as the contamination is concerned. With the fraction
of point sources misclassified as extended by the SASS being about
half as big as the one found for VTP (see Fig.~\ref{vtp_extent}) the
contamination of a SASS X-ray selected cluster sample can be expected
to lie somewhere in the range of values delineated by the dot-dashed
lines in Fig.~\ref{vtp_xselect}.

\begin{figure*}
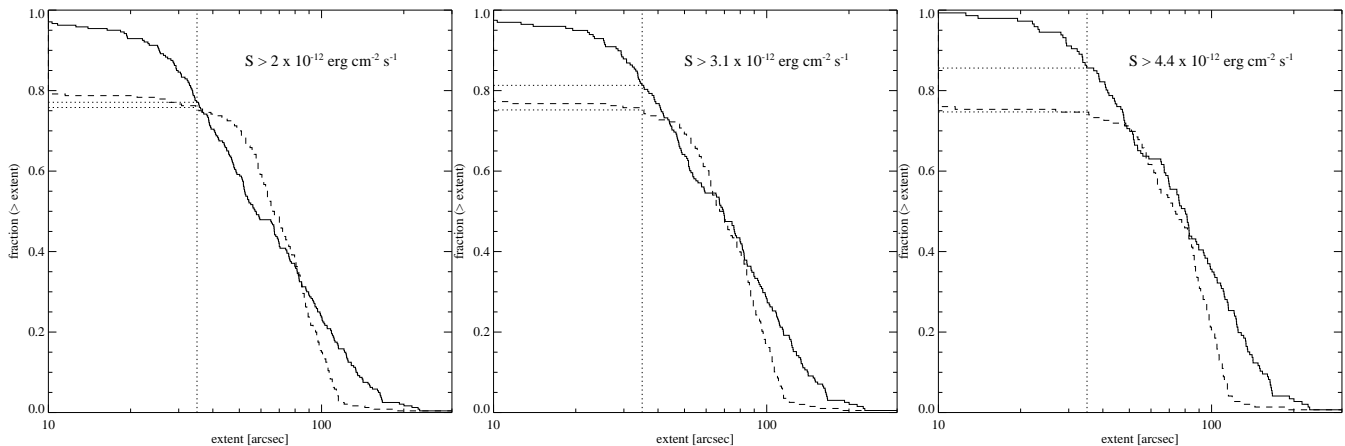

  \parbox{0.33\textwidth}{
  \epsfxsize=0.37\textwidth
  \hspace*{-6mm}
  \epsffile{bcs_vtp_sass_extent_1.epsf}}
  \parbox{0.33\textwidth}{
  \epsfxsize=0.37\textwidth
  \hspace*{-6mm}
  \epsffile{bcs_vtp_sass_extent_2.epsf}}
  \parbox{0.33\textwidth}{
  \epsfxsize=0.37\textwidth
  \hspace*{-6mm}
  \epsffile{bcs_vtp_sass_extent_3.epsf}}
  \caption[]{The cumulative extent distribution of the
           BCS clusters detected by both the SASS and VTP for various
           flux limits. The dotted lines mark the fraction of clusters
           and non-cluster sources classified as extended by each
           algorithm when the extent threshold is set to 35 arcsec. In
           each panel, the solid line shows the extent distribution
           returned by VTP while the dashed curves represent the
           distribution of SASS extents. The chosen flux cuts
           correspond to the lowest possible RASS flux limit of
           completeness and the BCS flux limits of 80 and 90 per cent
           completeness.}
           \label{vtp_sass_extent}
\end{figure*}

Unfortunately, the SASS not only detects fewer point sources as
extended, but is also much less efficient in detecting the extent of
truly extended sources, in particular when the emission is of low
surface brightness thus providing little contrast with the
background. This flaw goes hand in hand with the fact that the SASS
heavily underestimates the flux from extended emission (cf.\ Fig.~22
of EVB). Fig.~\ref{vtp_sass_extent} compares the extent values
returned by the SASS and VTP for those BCS clusters that have been
detected by both algorithms. (For obvious reasons we limit this
comparison to detections of Abell and Zwicky clusters and do not
include sources initially selected because of their SASS extent.)
Note that this comparison is favourable to the SASS as it does not
reflect a significant number of sources in the respective flux ranges
that the SASS has missed altogether (see Section~\ref{serdet}).  We
find that, at the lowest possible RASS flux limit of completeness,
both algorithms are about 75 per cent efficient in recognizing
clusters as extended X-ray sources when a threshold of 35 arcsec is
chosen for the extent parameter. As mentioned earlier in
Section~\ref{sass_sample}, this is the best possible extent threshold
for the SASS, as the fraction of SASS-extended BCS clusters does not
increase significantly when a lower value is chosen (see also
Fig.~\ref{exexl}). As, for this largest of the three flux-limited
samples of common detections depicted in Fig.~\ref{vtp_sass_extent},
VTP is equally efficient at the same numerical extent value, we will
use 35 arcsec as the common threshold for the extent parameters
returned by either algorithm in the comparison at the end of this
section. Note, however, that a lower threshold leads to a
significantly increased fraction of VTP-extended clusters while the
SASS efficiency remains essentially constant. Also, the maximal
completeness achievable with a SASS-extent-selected sample does not
increase any further, i.e., beyond 75 per cent, when the flux limit of
the sample is raised, which is mainly due to the poor correlation
between the SASS-detected fluxes and the true cluster fluxes (EVB).
As far as VTP goes, the flux limit does, however, have an effect on
the fraction of clusters recognized as extended: Of the 198 (145) BCS
clusters at fluxes greater than $3.1\times 10^{-12}$ erg s$^{-2}$
cm$^{-2}$ ($4.4\times 10^{-12}$ erg s$^{-2}$ cm$^{-2}$) detected by
both the SASS and VTP (again excluding the ones selected only because
of their SASS extent), VTP finds 81 (86) per cent to be significantly
extended in the X-ray.

As mentioned before, the resulting 25 per cent incompleteness of a
cluster sample selected exclusively by SASS extent at essentially all
flux levels is only the component due to unrecognized extended
sources. An {\em additional}\/ incompleteness of more than 20 per cent
(at fluxes of $3.1\times 10^{-12}$ erg s$^{-2}$ cm$^{-2}$) is caused
by the fact that 1 out of 5 clusters is missed altogether at the SASS
count rate cut of 0.1 count s$^{-1}$ that our study started from. This
number can certainly be reduced by including fainter SASS detections
to begin with -- however, all these additional sources need to be
re-processed by an algorithm that allows the true fluxes to be
determined. With the ratio between true and SASS-detected count rates
exceeding 2.3 (3.0) for more than 25 (11) per cent of the SASS
detections of clusters with true fluxes higher than $3.1\times
10^{-12}$ erg s$^{-2}$ cm$^{-2}$, this implies that all SASS
detections at least down to count rates of 0.046 count s$^{-1}$ would
need to be re-evaluated. This, however, is the SASS detection limit in
the RASS, i.e., even if {\em all}\/ extended SASS sources were
re-processed by VTP or some other algorithm, this would not even
ensure {\em detection}\/ completeness at $3.1\times 10^{-12}$ erg
s$^{-2}$ cm$^{-2}$. Also, the 25 per cent incompleteness due to
unrecognized extent would of course remain.

Given these difficulties it is not obvious how a statistically
complete and entirely X-ray selected cluster sample can ever be
compiled from SASS data at fluxes below the BCS flux limit of 90 per
cent completeness, without essentially re-processing the entire RASS
raw data with an additional algorithm that allows the SASS' erroneous
flux and extent determinations to be corrected.

If, however, priority is given to low contamination rather than
completeness, a SASS selected cluster sample might be slightly
preferred over one based on VTP detections. The resulting sample will
be at best 70 per cent complete but the contamination by point sources
would be no higher than about 40 per cent, compared to some 55 per
cent for a VTP based sample. The latter would, however, be 85 to 90
per cent complete.

\begin{figure*}
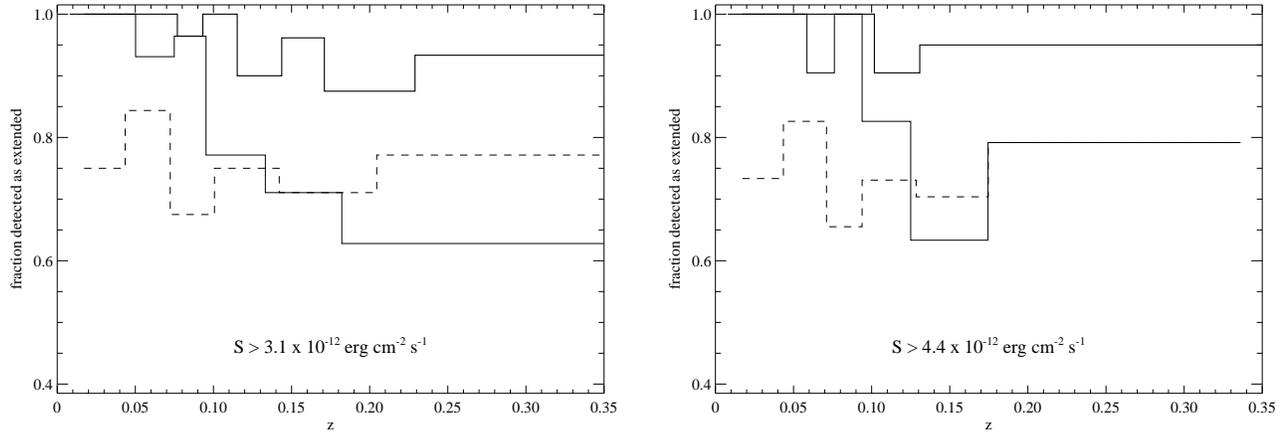

  \parbox{0.5\textwidth}{
  \epsfxsize=0.49\textwidth
  \epsffile{bcs_extent_z_1.epsf}}
  \parbox{0.5\textwidth}{
  \epsfxsize=0.49\textwidth
  \epsffile{bcs_extent_z_2.epsf}}
  \caption[]{The fraction of common (SASS and VTP detected) Abell and
           Zwicky clusters in the BCS that feature an X-ray source
           extent of more than 35 arcsec, at the flux limits of the 80
           and 90 per cent complete BCS subsamples (left and right
           panel, respectively).  In each panel, the thin solid line
           refers to VTP while the dashed one shows the SASS
           performance. The thick solid lines represent the fraction
           of {\em all}\/ BCS clusters at the respective flux limit
           that have been classified as extended by either or both of
           the two algorithms, i.e. -- in addition to Abell and Zwicky
           clusters -- it also takes into account the systems
           originally selected on the ground of their SASS extent. In
           the left-hand plot, each redshift interval but the first in
           the various histograms contains 27 clusters, while in the
           right-hand plot the respective number is 19.}
           \label{bcs_extent_z}
\end{figure*}

Turning again to the flux limited BCS subsamples, we confirm the
results of EVB on the redshift dependence of the cluster fraction
detected as extended by the SASS or VTP. Fig.~\ref{bcs_extent_z} shows
the fraction of extended sources in the sample of BCS clusters
detected by both algorithms as a function of redshift (again, clusters
originally included solely on the grounds of their SASS extent have
been excluded). As EVB did for their ACO cluster sample, we find the
SASS to perform at the about 75 per cent level, independent of
redshift and flux limit.  The fraction of VTP-extended BCS clusters is
similar (about 70 per cent for either flux limit) at redshifts greater
than 0.1, but rises steeply to more than 95 per cent for more nearby
clusters.

When all 283 (203) clusters of the 80 (90) per cent complete,
flux-limited BCS are considered, the total fraction of clusters
classified as extended by either or both of the two algorithms
reaches an impressive 96 (97) per cent, which varies only little
with redshift (bold solid lines in Fig.~\ref{bcs_extent_z}).

\subsection{Summary}

To summarize, we confirm that cluster selection by X-ray source extent
and hardness ratio works. However, due to the (different)
imperfections of either of the two detection algorithms used in the
present study (SASS and VTP), the compilation of {\em statistically
useful}\/ cluster samples requires a massive amount of work.  With
VTP, the price to be paid for a highly flux complete sample (85 to 90
per cent) is a considerable point source contamination of some 55 per
cent (35 per cent if an improved version of VTP is used). These point
sources need to be removed from the sample by visual inspection of the
X-ray detections and potential optical counterparts. With the SASS,
the contamination by point sources is slightly lower (typically 40 per
cent) but the highest achievable completeness is no more than about 70
per cent. Due to the biases introduced by such a low completeness, the
statistical usefulness of such a purely SASS-based cluster sample must remain
doubtful.

\end{document}